\documentclass[journal=jacsat,manuscript=article]{achemso}

\usepackage[version=3]{mhchem} \usepackage{soul}
\usepackage{setspace}
\usepackage{hyperref}
\usepackage{graphicx}%
\usepackage{multirow}%
\usepackage{amsmath,amssymb,amsfonts}%
\usepackage{amsthm}%
\usepackage{mathrsfs}%
\usepackage[title]{appendix}%
\usepackage{xcolor}%
\usepackage{textcomp}%
\usepackage{manyfoot}%
\usepackage{booktabs}%
\usepackage{algorithm}%
\usepackage{algorithmicx}%
\usepackage{algpseudocode}%
\usepackage{listings}%

\usepackage[normalem]{ulem}

\hypersetup{
    colorlinks=true,
    linkcolor=blue,
    filecolor=magenta,      
    urlcolor=cyan,
    pdftitle={Overleaf Example},
    pdfpagemode=FullScreen,
    }

\theoremstyle{thmstyleone}%

\theoremstyle{thmstyletwo}%

\theoremstyle{thmstylethree}%

\makeatletter
\let\l@addto@macro\relax
\makeatother
\usepackage[fontsize=10pt]{scrextend}

\usepackage{graphicx}%
\usepackage{multirow}%
\usepackage{amsmath,amssymb,amsfonts}%
\usepackage{amsthm}%
\usepackage{mathrsfs}%
\usepackage[title]{appendix}%
\usepackage{xcolor}%
\usepackage{textcomp}%
\usepackage{manyfoot}%
\usepackage{algorithm}%
\usepackage{algorithmicx}%
\usepackage{algpseudocode}%
\usepackage{listings}%

\author{Alessandro Tuniz}
\affiliation[Commonwealth Scientific and Industrial Research Organisation (CSIRO), Lindfield,
NSW, Australia]{Commonwealth Scientific and Industrial Research Organisation (CSIRO), Lindfield,
NSW, Australia}
\alsoaffiliation[Institute of Photonics and Optical Science (IPOS), School of Physics, The University of Sydney, Sydney, NSW 2006, Australia]
{Institute of Photonics and Optical Science (IPOS), School of Physics, The University of Sydney, Sydney, NSW 2006, Australia}
\email{alessandro.tuniz@csiro.au}
\author{Sabrina Garattoni}
\affiliation[Dipartimento di Fisica, Politecnico di Milano, P.zza Leonardo da Vinci 32, Milan, 20133 Italy]
{Dipartimento di Fisica, Politecnico di Milano, P.zza Leonardo da Vinci 32, Milan, 20133 Italy}
\alsoaffiliation[Institute of Photonics and Optical Science (IPOS), School of Physics, The University of Sydney, Sydney, NSW 2006, Australia]
{Institute of Photonics and Optical Science (IPOS), School of Physics, The University of Sydney, Sydney, NSW 2006, Australia}
\author{Han-Hao Cheng}
\affiliation[Centre for Microscopy and Microanalysis, The University of Queensland, Brisbane, QLD, Australia]
{Centre for Microscopy and Microanalysis, The University of Queensland, Brisbane, QLD, Australia}

\author{Giuseppe Della Valle}
\affiliation[Dipartimento di Fisica, Politecnico di Milano, P.zza Leonardo da Vinci 32, Milan, 20133 Italy]
{Dipartimento di Fisica, Politecnico di Milano, P.zza Leonardo da Vinci 32, Milan, 20133 Italy}

\title[Article Title]{Directional coupling to a $\lambda/5000$ nano-waveguide}

\begin{document}

\maketitle

\clearpage

\begin{abstract}
{Silicon-based micro-devices are considered promising candidates for consolidating several terahertz technologies into a common and practical platform. The practicality stems from the relatively low loss, device compactness, ease of fabrication, and wide range of available passive and active functionalities. Nevertheless, typical device footprints are limited by diffraction to several hundreds of micrometers, which hinders emerging nanoscale applications of terahertz frequencies. While metallic gap modes provide nanoscale terahertz confinement, efficiently coupling to them is difficult. Here we present and experimentally demonstrate a strategy for efficiently interfacing sub-terahertz radiation ($\lambda=1\,{\rm mm}$) to a waveguide formed by a nanogap, etched in a gold film, that is 200\,{\rm nm} ($\lambda/5000$) wide and up to 4.5\,mm long. The design principle relies on phase matching dielectric and nanogap waveguide modes, resulting in efficient directional coupling between them when placed side-by-side. Broadband far field terahertz transmission experiments through the dielectric waveguide reveal a transmission dip near the designed wavelength due to resonant coupling. Near field measurements on the surface of the gold layer confirm that such a dip is accompanied by a transfer of power to the nanogap, with an estimated coupling efficiency of $\sim 10\%$. Our approach provides a pathway for efficiently interfacing millimeter-wave and near-infrared photonic circuits, providing controlled and tailored nanoscale terahertz confinement, with important implications for on-chip nanospectroscopy, telecommunications, and quantum technologies.}
\end{abstract}


\newpage 



\clearpage 

\section{Introduction}

Terahertz radiation encompasses frequencies in the 0.1-10\,THz range -- corresponding to wavelengths of 30\,mm--3\,mm -- and is increasingly harnessed for numerous different, far-reaching, and cross-disciplinary applications. Examples include bio-sensing~\cite{markelz2022perspective, smolyanskaya2018terahertz} and spectroscopy~\cite{Seo2022}, as well as space~\cite{siegel2007thz} and ground~\cite{kleine2011review} wireless communication. Historically, the development of efficient terahertz sources has been hindered by fundamental limitations that make both electrically-driven and optically-inspired approaches inefficient~\cite{sizov2010thz}. Nevertheless, recent decades have been marked by rapid technological progress, and the overall consensus is that the terahertz gap is rapidly being bridged~\cite{pang2022bridging}. 

As terahertz sources and detectors have become increasingly available, the focus has thus started to shift on developing practical devices and experimental techniques for harnessing terahertz radiation effectively. On the one hand, it is important to develop unified and reliable protocols to aid reproducibility in the context of bio-spectroscopy~\cite{markelz2022perspective}; on the other, it is becoming increasingly important to access a library of chip-scale integrated components that can manipulate terahertz radiation without using bulky free space optics~\cite{headland2020unclad, xu2022wired}. Indeed, a plethora of applications emerge when terahertz- and nano-technology converge, particularly in biological analysis, electronic and photonic devices, imaging, spectroscopy, and sensing~\cite{leitenstorfer20232023}. For example, nanoscale and macroscale properties can differ significantly~\cite{lawler2020convergence, rosei2004nanostructured}, warranting practical and reliable platforms to study such regimes. In this context, the fundamental challenge is that the wavelength of terahertz radiation is at least three
orders of magnitude larger than those needed in the realms of nanotechnology. With conventional techniques, addressing individual nanostructures terahertz at frequencies is impossible due to the diffraction limit~\cite{chen2003terahertz}; this can be overcome with terahertz scanning near-field optical microscopy, which can achieve microscale resolution over large areas~\cite{wittmann2023assessment, tuniz2023subwavelength}, and  nanoscale resolution (i.e., nanoscopy~\cite{guo2024terahertz}) over smaller areas~\cite{cocker2021nanoscale}, even in aqueous environments~\cite{kaltenecker2021infrared}. Terahertz nanotechnology has also benefited from the development of increasingly efficient terahertz sources~\cite{park2012enhancement, PetersAOM2024}, detectors~\cite{vitiello2012room}, and modulators~\cite{degl2018all}, by enhancing light-matter interactions in small volumes. We point the reader to Refs.~\cite{lawler2020convergence, guo2024terahertz} and references therein for an overview of terahertz nanotechnology.

\begin{figure*}[t!]
\centering
\includegraphics[width=\textwidth]{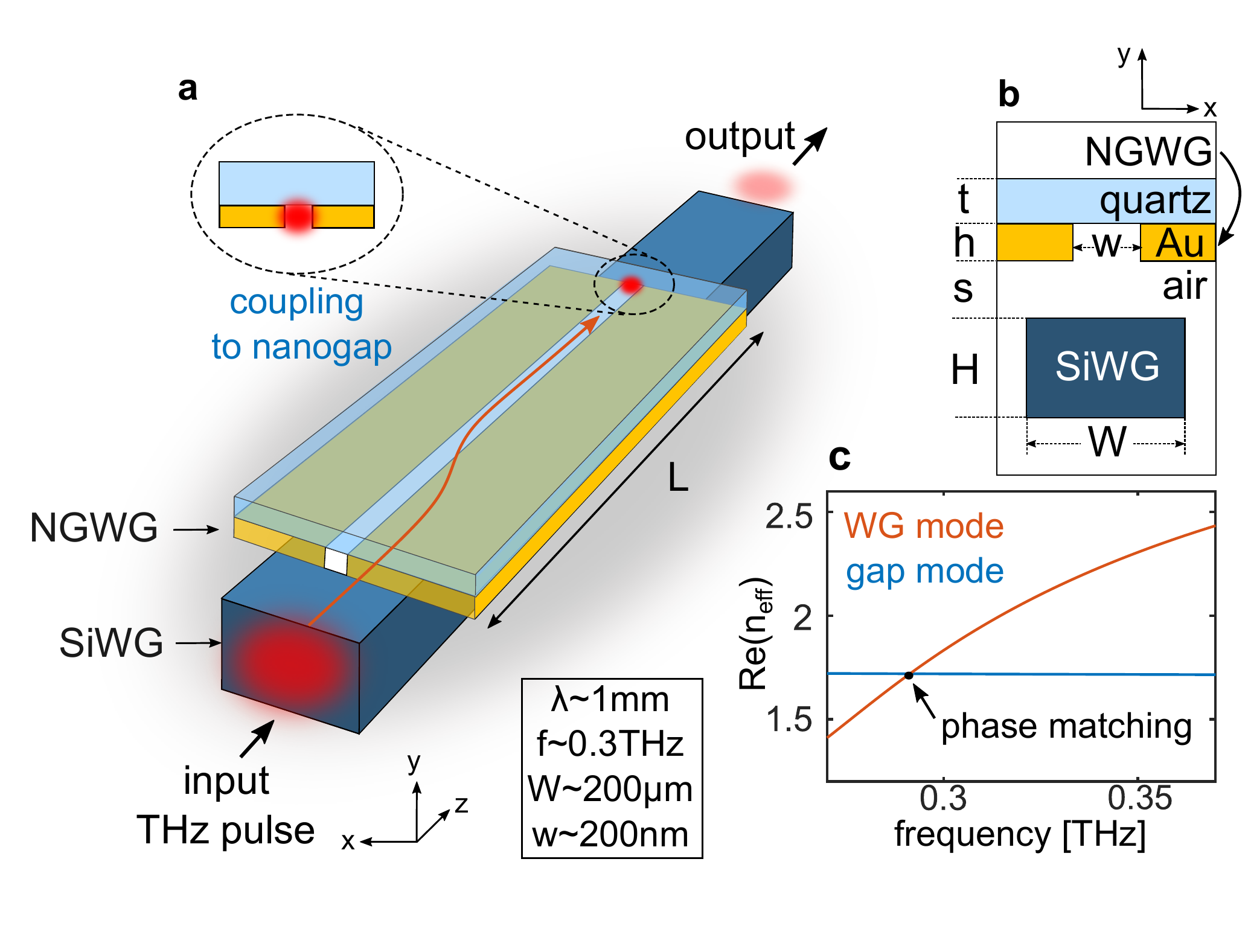}
\caption{Concept schematic of the terahertz directional nano-coupler presented here. (a) An $x$-polarized broadband terahertz pulse propagates in $z$ through a substrateless silicon waveguide (SiWG). At $\lambda \sim 1\,{\rm mm}$, it efficiently couples to the metallic gap mode over a length $L$, due to phase matching with the fundamental mode of the nanogap waveguide (NGWG) above it. (b) Cross section schematic of the coupling region in the $xy$ plane, showing geometry and material distribution (light blue: quartz; dark blue: silicon; yellow: gold; white: air). Here, $t=50\,\mu{\rm m}$, $h=100\,{\rm nm}$, $H=250\,\mu{\rm m}$, $W=190\,\mu{\rm m}$, $w=200\,{\rm nm}$, and $s$ is varied. Note that the schematic is not to scale to simplify labeling, and that waveguide and gap widths differ by three orders of magnitude:  $w\sim\lambda/5000$. (c) Calculated dispersion of $\Re e(n_{\rm eff})$ of the fundamental modes of the individual SiWG (red) and NGWG (blue), as a function of frequency, highlighting the phase matching point near 0.3\,THz.}
\label{fig:1}
\end{figure*}

In parallel, there has been growing interest to develop a general-purpose integrated terahertz platform. For example, in Ref.~\cite{headland2023terahertz} Headland \emph{et al.} argue that all-silicon substrateless waveguides -- whose cross section is of order $~200\,\times 200{\rm {\mu}m}^2$, and whose length is of the order of the centimeters -- are one of the more promising candidates, because of their intrinsic compactness, ease-of-design, and wide variety of accessible functionalities. Nevertheless, the lateral field confinement in such platforms is intrinsically limited by diffraction to $\sim 100{\rm {\mu}m}$ at terahertz frequencies, hindering nanoscale applications. The only practical way to confine radiation to deep-subwavelength scales is by using metallic components~\cite{han2015spoof} which, in the simplest configuration, can be assembled into a nano-mirror cavity that confines linearly polarized radiation inside a nanogap, giving rise to a nanoscale-bound and propagating terahertz mode~\cite{tsiatmas2012low}. This is commonly referred to as the metal/dielectric/metal (MDM) configuration although similar properties can be achieved by nanowires using radial polarization~\cite{yang2010theory}. This confinement is typically accompanied by Ohmic losses, which constitute an important limitation to practical applications~\cite{tsiatmas2012low}. In addition, the small mode area makes direct coupling to such modes from free space challenging. At terahertz frequencies, this problem is further exacerbated by the enormous difference in size and field  distribution between dielectric- and MDM modes, which in the present context requires a millimeter-to-nanometer mode converter. As a result, most integrated optical-to-terahertz generation and detection schemes use the chip component to support optical waveguiding, which in turn interfaces with an external (free space) terahertz fields~\cite{salamin2019compact} -- we refer to Refs.~\cite{sengupta2018terahertz, rajabali2023present} for recent reviews of hybrid integrated terahertz photonic devices. Furthermore, most schemes that use metallic nanogaps, e.g., for sensing~\cite{park2015nanogap, park2017sensing, kim2018colossal} or terahertz detection~\cite{lee2015highly}, do so in a planar (metasurface) configuration of deep subwavelength thickness and under free space illumination that is perpendicular to the surface~\cite{bahk2019large}, with an intrinsically small overlap between the incoming diffraction-limited terahertz beam and the nanoscale gaps. In contrast, this work considers the case where the direction of propagation in $z$ is parallel to the surface, so that  the nanogap acts as a waveguide.

In order to achieve an all-integrated  platform where \emph{both} optical and terahertz signals propagate on a photonic chip over wavelength scales, it is first necessary to efficiently transfer power from a guided-wave terahertz signal to an area of $\sim 200 \times 200 \,{\rm nm}^2$, comparable to that of typical photonic waveguides. The most direct approach is arguably end-fire coupling from the gap waveguide edge~\cite{stegeman1983excitation} -- however, in the present configuration the huge mismatch between the guided mode and a diffraction limited free space beam leads to low coupling efficiencies.
Another approach adapted at near-infrared frequencies, which is also suited for photonic integration, is  directional- or adiabatic- coupling~\cite{taras2021shortcuts}, whose underlying formalism relies on matching the propagation constants of two adjacent modes~\cite{tuniz2024coupled}, and is agnostic to the physical dimensions of the waveguides. Most recently for example, a millimeter-wave mode converter was shown to efficiently transfer signals from a silicon waveguide to an antenna of $\sim 10\,\mu{\rm m}$ lateral dimensions placed directly on top~\cite{yu2019efficient}. More broadly, several nanoscale mode converters on the basis of directional~\cite{delacour2010efficient} and adiabatic~\cite{nielsen2017giant} coupling have been developed in silicon-on-insulator substrates at infrared wavelengths. The principle of operation relies on the fact that the effective index of nanoscale metal waveguides (e.g., metal nano-wires or nano-gaps) can be significantly above that of surrounding medium when the waveguide dimensions are comparable to the skin depth of the metal at that frequency~\cite{novotny2012principles, seo2009terahertz}. Therefore, even metal waveguides in air can be phase matched with high-index dielectric waveguides; at terahertz frequencies, the skin depth of most metals is $\sim 100\,{\rm nm}$~\cite{he2009investigation}, which immediately suggests the potential for phase-matching between nanometer metal waveguides and wavelength-scale dielectric waveguides.

A concept schematic of the device layout and operating principles are shown in Fig.~\ref{fig:1}(a). The device is composed of a millimeter-scale silicon waveguide (SiWG) which is adjacent to a metallic nanogap waveguide (NGWG), and whose lateral dimensions differ by three orders of magnitude. The NGWG rests on a quartz substrate, with length $L = 3.7-4.5\,{\rm mm}$. 
The input is a broadband terahertz pulse (frequency: 0.2--3\,THz) propagating in the silicon waveguide as per Fig.~\ref{fig:1}(a).  Note that both the SiWG and the NGWG substrate are suspended in air, and that they can be easily and independently handled with micrometer stages. A schematic of the cross section  of the device is shown in Fig.~\ref{fig:1}(b). The SiWG has a height $H=250\,\mu{\rm m}$ and width $W=190\,\mu{\rm m}$; the NGWG has a width $w=200\,{\rm nm}$ height $h=100\,{\rm nm}$, supported by a quartz substrate of thickness $t=50\,\mu{\rm m}$. The edge-to-edge separation between the waveguides is $s$, which here lies in the range 10--150\,$\mu{\rm m}$. 

We first consider the two waveguides separately (i.e., assuming $s=\infty$). The calculated dispersion of the real part of the effective index $n_{\rm eff}$ of the fundamental $x$-polarized suspended SiWG mode is shown as a red line in Fig.~\ref{fig:1}(c), assuming a constant refractive index of 3.42 for silicon~\cite{dai2004terahertz} and neglecting losses for simplicity. Note the rise in effective index with frequency, due to an increase in the fraction of the mode field in high-index silicon. The corresponding $\Re e(n_{\rm eff})$ of the NGWG mode is shown as a blue line, which is nominally independent of frequency in this range, considering a lossless refractive index of 2.09 for quartz~\cite{davies2018temperature}, and taking a Drude model for gold~\cite{rakic1998optical}. Under these conditions, the SiWG and NGWG mode phase match at 0.29\,THz, and are expected to couple as they are brought closer, leading to the emergence of two supermodes that produce efficient power transfer between waveguides via directional coupling. We now discuss this in more detail for the present geometry.

\begin{figure*}[t!]
\centering
\includegraphics[width=\textwidth]{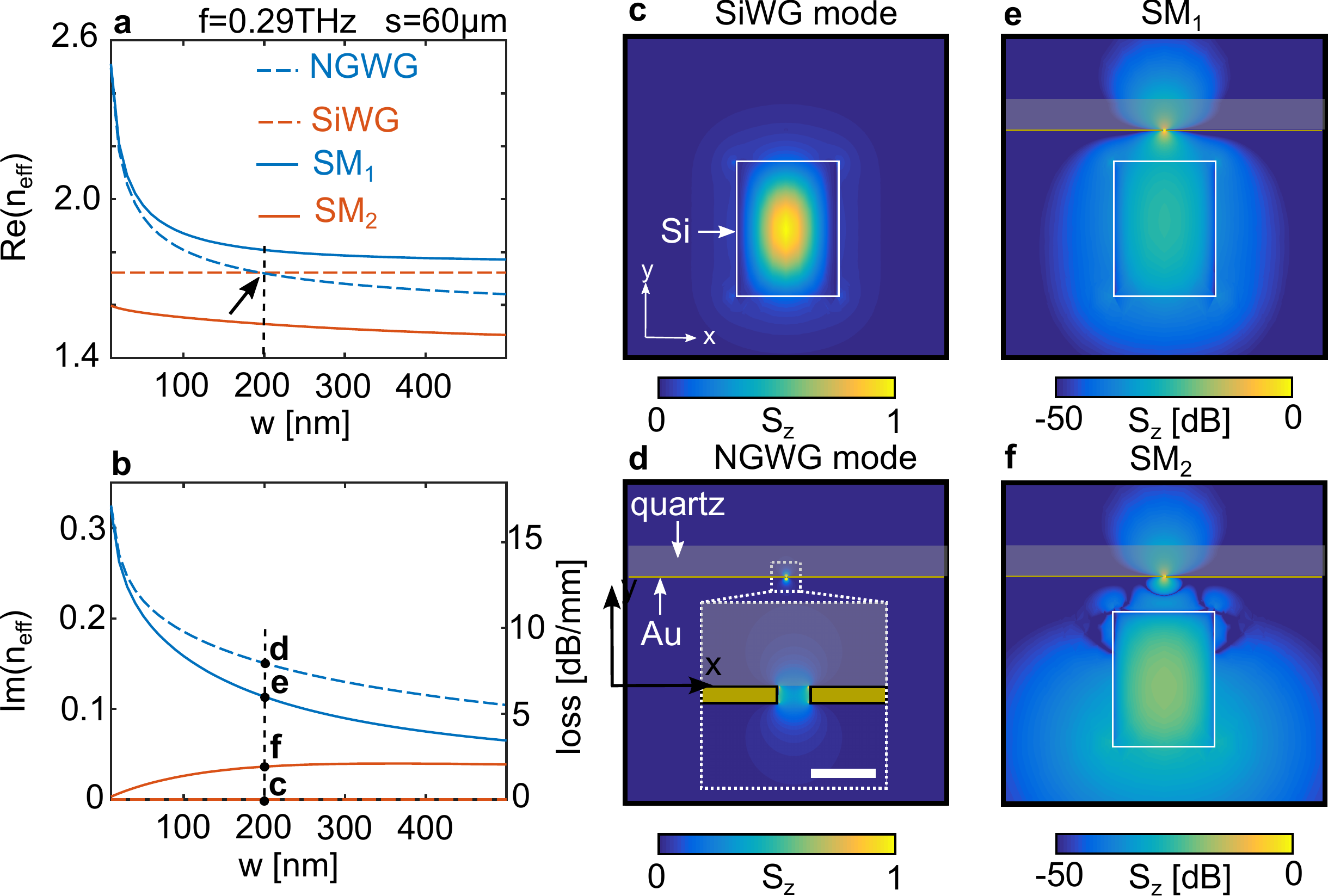}
\caption{(a) Real part of the effective index of relevant modes as a function of the gap width $w$, at $f=0.29\,{\rm THz}$ and $s=60\,\mu{\rm m}$, with all other parameters as per the Fig.~\ref{fig:1} caption. The dielectric WG (SiWG) remains unchanged with $w$ (red dashed line). The effective index of the nanogap mode increases with decreasing $w$ (blue dashed line), and phase matches with the WG mode at $w=200\,{\rm nm}$. The modes hybridize when the waveguides are adjacent, resulting in two supermodes (SMs, solid lines), whose effective indices split at the phase-matching point. (b) Corresponding imaginary part of the effective index. (c) and (d) show the colourmap of the WG and gap modes ($z$-component of the Poynting vector), with a detail of the white dashed region in the inset of panel (d). (e) and (f) respectively show the hybrid SM$_1$ and SM$_2$ that emerge when the waveguides couple. A logarithmic scale is used due to the large relative field intensity in the nanogap. Window size in (c)--(f): $600\,\mu{\rm m} \times 600\,\mu{\rm m}$. Scale bar in (d) inset: 400\,nm.}
\label{fig:2}
\end{figure*}

\section{Results}

Before discussing the power transfer from the SiWG to the NGWG, we first consider the modes supported by the coupled waveguides during propagation along the device length $L$ as per Fig.~\ref{fig:1}. The most intriguing aspect to analyse, in first instance, is how the phase-matching between the SiWG and the NGWG emerges, and how this leads to mode hybridization. The blue dashed lines in Fig.~\ref{fig:2}(a) and~\ref{fig:2}(b) shows the result of 2D finite element method (FEM) simulations with a commercial tool (COMSOL Multiphysics) of the real- and imaginary- parts of the effective index of the fundamental mode of the isolated NGWG as a function of $w$ at a constant frequency of 0.29\,THz -- see Methods for details of the simulation and material parameters used. Note in Fig.~\ref{fig:2}(a) the sharp increase in both real- and imaginary- parts of $n_{\rm eff}$ as $w$ is decreased, with $n_{\rm eff}$ approaching the effective index of the quartz slab mode as $w$ increases. The right axis of Fig.~\ref{fig:2}(b) plots the associated loss, which is $5\,{\rm dB/ mm}$, indicating  that millimeter-scale propagation of this NGWG mode is supported. The $x$-polarized mode effective index of the isolated SiWG is fixed (being $w$ independent), purely real (i.e., lossless), and marked by the horizontal red dashed line in Fig.~\ref{fig:2}(a). The Poynting vector associated with this mode is shown in Fig.~\ref{fig:2}(c), whereas that of the isolated NGWG mode is shown in Fig.~\ref{fig:2}(d). Note that the latter possesses a vastly reduced mode volume compared to the SiWG mode. A zoomed-in view of this mode is shown in the inset of Fig.~\ref{fig:2}(d) for $w=200\,{\rm nm}$. If the two waveguides are placed side by side (e.g., $s=60\,\mu{\rm m}$), then the modes hybridize, giving rise to so-called supermodes (SMs), which constitute the modes of the two-waveguide system. According to coupled mode theory, under the weak coupling approximation, SMs are even- and odd- superpositions of the isolated modes at the phase matching point~\cite{huang1994coupled}. Here, the modes are strongly coupled and their propagation constants and field distributions must be calculated numerically. The solid lines in Fig.~\ref{fig:2}(a) and~\ref{fig:2}(b) respectively show the real- and imaginary- parts of the effective index of the SMs as a function of $w$ at 0.29\,THz. Note that the $\Re e(n_{\rm eff})$ now anti-cross at the phase matching point, and that both modes are lossy (i.e., $\Im m(n_{\rm eff})>0$ for both SMs). The colormaps of the associated SMs for $w=200\,{\rm nm}$ (Fig.~\ref{fig:2}(e) and~\ref{fig:2}(f), respectively), clearly show modal hybridization.

Because the propagation constants of the participating isolated modes are complex valued, the coupler is non-Hermitian~\cite{feng2017non}, and the resulting mode hybridization subtly depends on the interplay of coupling strength and loss -- see, for example, Ref.~\cite{miri2019exceptional} for a review of non-Hermitian photonics, and Ref.~\cite{tuniz2022influence} for a non-Hermitian perspective on plasmonic couplers. For ease of discussion, here we only consider how the SMs of the two-waveguide system depend on frequency and separation, and device length in the present context -- which we will subsequently experimentally access --  and how this is expected to influence the transmission through the dielectric waveguide and the associated power transfer to the NGWG.

\subsection{Simulations}

\begin{figure*}[t!]
\centering
\includegraphics[width=\textwidth]{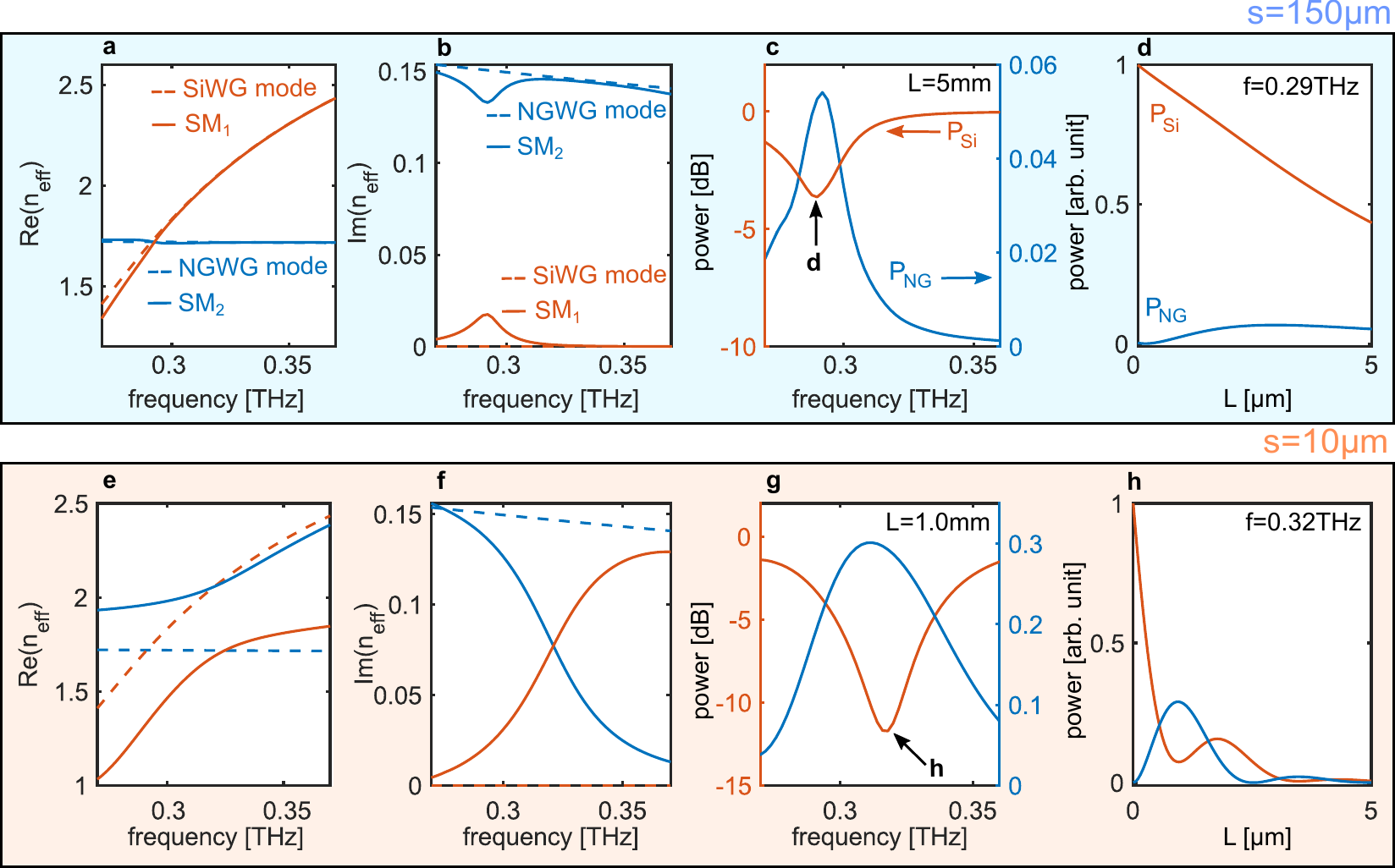}
\caption{Overview of the effective index dispersion, transmission spectra, and power exchange properties for different separations $s$. (a) For larger separations ($s=150\,\mu{\rm m}$, blue box) the real part of the effective index for both isolated and coupled WG/gap modes cross, and (b) the imaginary parts anti-cross. (c) Associated calculated power in the dielectric waveguide $P_{\rm Si}$ as a function of frequency for $L=5\,{\rm mm}$ (red line, left axis on dB scale), showing resonant transmission dip accompanied by a corresponding transmission peak in the nano-gap waveguide power $P_{\rm NG}$ (blue line, right axis). At the resonance frequency (0.29\,THz, black arrow), (d) shows that the $P_{\rm Si}$ monotonically decreases with increasing length $L$, with $P_{\rm NG}$ plateauing at $\sim 0.05$, decreasing at longer lengths due to losses. Bringing the waveguides closer ($s=10\,\mu{\rm m}$, green box) leads to (e)  an anti-crossing in the the real parts of the SM dispersions, (f) and a crossing of their imaginary parts, due to stronger coupling~\cite{tuniz2024coupled}. (g) The resulting transmission dip is sharper for shorter device lengths (here: $L=1.0$~mm) due to directional coupling (red lines, left axis). Correspondingly, more power is coupled to the plasmonic gap. Note that the $P_{\rm Si}$ minimum has shifted to 0.32\,THz (black arrow in (g)). (h) At this frequency, the power in the SiWG oscillates and decays (red line), whereas the relative power in the NGWG peaks at $L\sim1.0\,{\rm mm}$.}
\label{fig:3}
\end{figure*}

We are predominantly interested in the understanding of how energy is transferred between the dielectric and the metal gap waveguide under different configurations. Most commonly~\cite{bogaerts2018silicon}, complex coupler designs are understood and optimized via three-dimensional full-field numerical modelling, e.g., using finite element, or finite-difference time domain methods. However, such approaches are not practical in the present case, because the size of the gap is more than three orders of magnitude smaller than the propagation length, making memory requirements and calculation times prohibitive for full three-dimensional models~\cite{tuniz2024coupled}. Instead, we use the eigenmode expansion method, which relies on a combination of two-dimensional mode numerical calculations and analytical formulae~\cite{tuniz2024coupled, tuniz2016broadband} for propagation in $z$ -- see Methods for further details. We now discuss a few representative cases which elucidate the expected coupling between the waveguides under different conditions.

We consider all relevant eigenmodes as a function of frequency $f$ and separation $s$, taking $w=200\,{\rm nm}$, and all other parameters as in Fig.~\ref{fig:2}.  
The top row of Fig.~\ref{fig:3} shows a summary of the case where $s=150\,\mu{\rm m}$, i.e. at $\sim\lambda/6$ distance between the SiWG and the NGWG. Figure~\ref{fig:3}(a) and~\ref{fig:3}(b) detail the real and imaginary part of the isolated modes and SM dispersion, using the same notation as in Fig.~\ref{fig:2}. The isolated modes phase match, as per our plot in Fig.~\ref{fig:1}(c); however, contrary to our analysis in Fig.~\ref{fig:2}, Fig.~\ref{fig:3}(a) shows that the real parts of the SM effective indexes cross, precluding directional coupling, since the coupling length at the phase matching point, $L_c = \lambda / \Delta n_{\rm eff}$, turns out to be infinite~\cite{tuniz2019tuning}. This takes place because at the considered separation distance ($s\sim\lambda/6$) the coupling is too weak to produce eigenmode splitting~\cite{tuniz2024coupled}. Although these waveguides are very close ($s<\lambda/5$), the NGWG mode is confined close to the nanofilm (inset in Fig.~\ref{fig:2}(d)), and the effective index's imaginary part is relatively large ($\sim 10\%$ of its real part), precluding the SMs' avoided crossing. The associated $\Im m (n_{\rm eff})$, shown in Fig.~\ref{fig:3}(b), anti-cross. As expected~\cite{tuniz2019tuning}, the loss of SM$_1$ increases near the phase matching point which, in turn, produces a drop in the overall transmission because  in this regime the fundamental mode of the incoming waveguide predominantly couples to SM$_1$~\cite{tuniz2022influence}. To confirm this, we calculate the normalized power in the dielectric waveguide $P_{\rm Si}$ as a function of frequency for $L=5\,{\rm mm}$ which is shown on a dB scale in Fig.~\ref{fig:3}(c) (red line, left axis), and presents a broad transmission dip at 0.29\,THz. The corresponding calculated normalized power in the nanogap waveguide $P_{\rm NG}$ is shown on a linear scale in Fig.~\ref{fig:3}(c) (blue line, right axis). Remarkably, we find that more than 5\% of the incoming power is predicted to coupled to the nanogap, despite having a lateral dimension of less than $\lambda/5000$. Phase matching, even at relatively large distances and without achieving eigenmode splitting, thus provides a pathway for accessing nanoscale mode areas from dielectric waveguides with percentage-level efficiency. Figure~\ref{fig:3}(d) shows $P_{\rm Si}$ and $P_{\rm NG}$ at the 0.29\,THz resonance as a function of $L$. Note that the power in SiWG monotonically decreases with $L$, whereas the power in NGWG reaches a plateau due to losses.

When the waveguides are brought closer and their modes couple more strongly, the power exchange properties of the nano-coupler are significantly modified. The bottom row of Fig.~\ref{fig:3} shows the plots corresponding to the top row but now taking $s=10\,\mu{\rm m}$. The solid lines in Fig.~\ref{fig:3}(e) and~\ref{fig:3}(f) respectively show the real and imaginary part of the SM dispersions. 
The coupling is now strong enough to produce eigenmode splitting~\cite{tuniz2024coupled}, and the real parts of the effective indices of the SMs anti-cross, while their imaginary parts cross. Note that the frequency of the SM anti-crossing point in Fig.~\ref{fig:3}(e) (i.e., where the difference in their $\Re e(n_{\rm eff})$ is minimum) occurs at a higher frequency than the phase matching point.  In this case, an accurate computation of the transmission spectra must take into accounting each mode's excitation and propagation along the device length $L$ via eigenmode expansion method~\cite{tuniz2016broadband} (see Methods section). 
The calculated $P_{\rm Si}$ as a function of frequency for $L=1\,\mu{\rm m}$ is shown on a dB scale in Fig.~\ref{fig:3}(g) (red line, left axis). Compared to the situation at larger separation (cf.~Fig.~\ref{fig:3}(c)), the transmission dip shifts to a higher frequency of 0.32\,THz. Furthermore, as a result of the short device length and the possibility of resonant interference between modes due to a splitting of the eigenmodes, the associated normalized $P_{\rm NG}$ increases to 0.3, peaking when $P_{\rm Si}$ is at a minimum as shown in Fig.~\ref{fig:3}(h) (blue line, right axis). Figure~\ref{fig:3}(h) shows $P_{\rm Si}$ and $P_{\rm NG}$ at 0.32\,THz, showing lossy oscillatory behaviour with increasing length, as expected for coupled lossy waveguides in this regime~\cite{tuniz2019tuning, tuniz2024coupled}.

\begin{figure*}[t!]
\centering
\includegraphics[width=\textwidth]{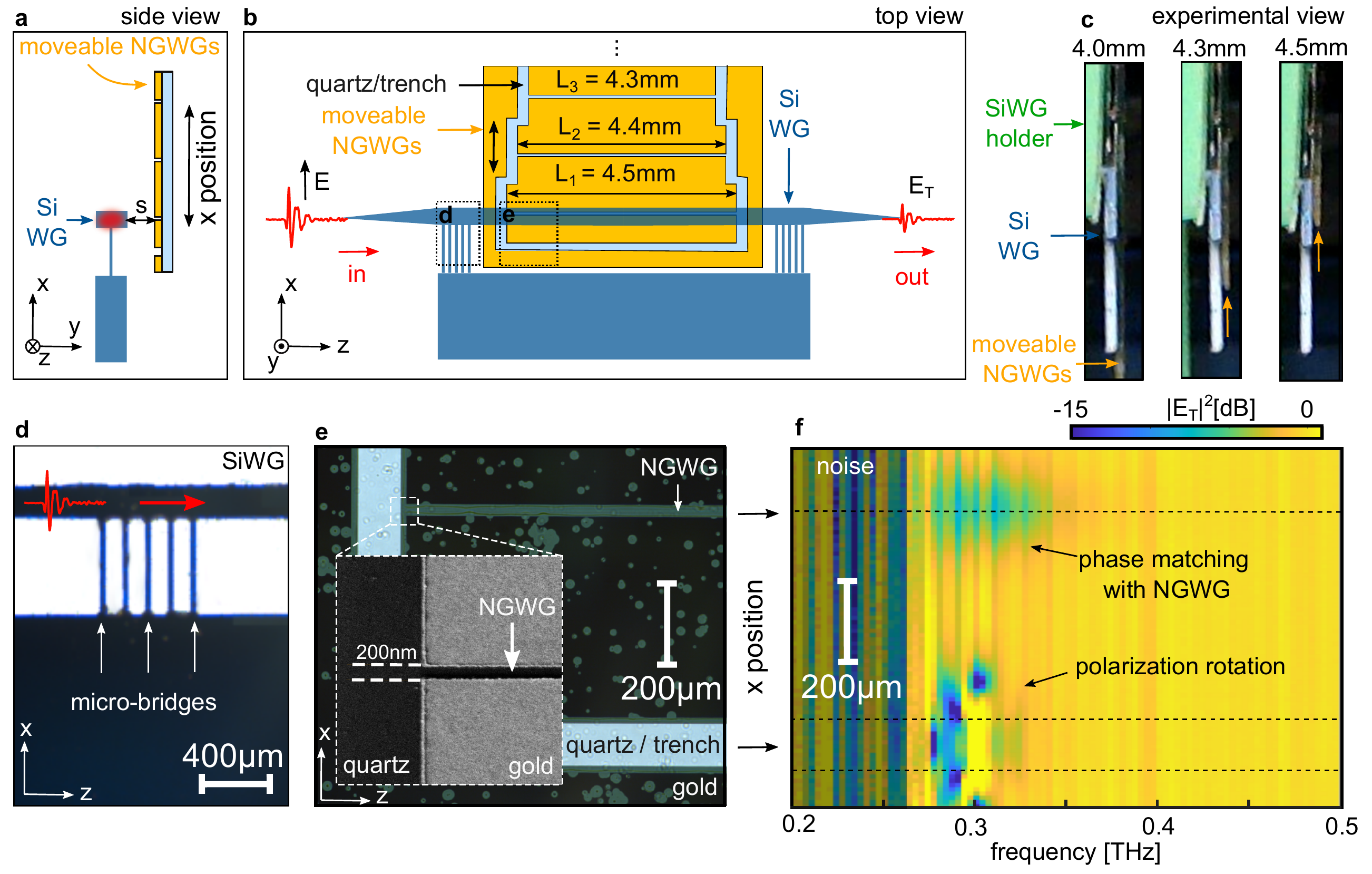}
\caption{Summary of the far field transmission experiments. (a) Side view of the experiments. An $x$-polarized THz pulse (red) is coupled to the fundamental mode of the substrateless silicon waveguide (SiWG, dark blue). A series of metal nanogap waveguides (yellow) on a quartz substrate (light blue) are moved in $x$ for a given separation $s$. (b) Top view of the experiment. The NGWGs vary in length (here: $L=3.7-4.5\,{\rm mm}$). We measure the transmitted intensity  $|E_T|^2$ of the THz pulse through the SiWG as a function of $x$. (c) Microscope images of the experiment for different values of $L$ as labelled. The green is a custom holder for the SiWG. Note the moving NGWGs for a fixed input condition. (d) Optical micrograph of the SiWG and microbridges. (e) Optical micrograph of the gold NGWGs deposited on quartz, surrounded by a $100\,\mu{\rm m}$ trench, included to facilitate visual inspection. Inset: scanning electron micrograph of the NGWG edge, showing the 200\,nm gap. (f) Measured colourmap of the transmitted intensity through the SiWG as a function of frequency and the $x$ position of the NGWG sample for a constant $s\sim50\,\mu{\rm m}$, covering both the trench and the 4.5\,mm NGWG sample.}
\label{fig:4}
\end{figure*}

\subsection{Far-field transmission experiments}
In a first set of experiments, we measure the transmission through a SiWG which is adjacent to a gold film containing a series of NGWGs of varying length. Figure~\ref{fig:4}(a) and~\ref{fig:4}(b) respectively show a side and top view schematic of the sample positioning and experimental setup. All materials are colour-coded as per Fig.~\ref{fig:1}(b). The SiWG layout is inspired by previous designs~\cite{akiki2020high, verstuyft2022short}: it is formed by a suspended silicon strand of 1\,cm length, which is connected to a silicon substrate (substrate area: $1\,{\rm cm}\times1\,{\rm cm}$) via subwavelength micro-bridges to minimize scattering of the fundamental mode upon propagation. An additional taper on each end of the waveguide (taper length: 1\,cm) facilitate coupling to- and from- free space. The NGWGs have lengths between 3.7-4.5\,mm. Each NGWG is separated in $x$ by 1\,mm. The NGWGs are etched on a 100\,nm gold film on a quartz substrate of $50\,\mu{\rm m}$ thickness. Each waveguide is surrounded by a trench of width $100\,\mu{\rm m}$ to facilitate alignment, highlighted in Fig.~\ref{fig:4}(b). We use a THz time domain spectroscopy (THz-TDS) system to measure the field transmitted by the waveguide. The electric field is polarized in $x$. Computer-controlled stages enable us to vary both $s$ and the $x$-position while the SiWG is kept still after coupling to its fundamental mode. 

Figure~\ref{fig:4}(c) shows an example experimental view obtained from an external microscope when the NGWGs are in near-contact with the SiWG, varying the $x$-position to align the SiWG with NGWGs of length $L$ as labelled. In detail, Fig.~\ref{fig:4}(d) shows an optical micrograph of the SiWG from the top, highlighting the microbridges and the direction of propagation of the pulse inside the waveguide. Figure~\ref{fig:4}(e) shows an optical micrograph top view of the NGWGs and trench, which exposes the underlying quartz substrate. We use scanning electron microscopy to confirm that all waveguides considered here are continuous for their entire length. For example, the inset of Fig.~\ref{fig:4}(e) shows the edge of the NGWG. To confirm that the SiWG mode is coupling to the NGWG mode, we first measure the terahertz pulse transmitted by the SiWG when it is adjacent to the gold/quartz substrate as a function of the $x$-position, 
varying $x$ over the region bounded by the micrograph of Fig.~\ref{fig:4}(e). Our experiment directly measures the terahertz pulse field emerging from the SiWG, so that its Fourier transform provides the transmitted amplitude and phase of the electric field at every frequency~\cite{jepsen2011terahertz}. Figure~\ref{fig:4}(f) shows a false colour plot of the resulting transmitted intensity as a function of frequency and $x$-position, where  $s\sim50\,\mu{\rm m}$ is kept constant. Note that we have aligned Fig.~\ref{fig:4}(e) and~\ref{fig:4}(f) in $x$ for better comparison, i.e., the vertical scale bars coincide. The dashed lines in Fig.~\ref{fig:4}(f) overlap with the edge of the trench and the center of the NGWG. We observe a series of sharp transmission dips near 0.3\,THz as we move the trench over the SiWG. Our 3D finite element calculations suggest that these dips are due to polarization rotation~\cite{caspers2013experimental, tuniz2020modular} -- see Figure~\ref{fig:rotator} in the Supporting Information -- which we measure because the detector records $x$-polarized fields. Analogous polarization rotation effects using gold nanostrips over a silicon waveguide have been studied in the near-infrared~\cite{kim2015polarization}. A detailed discussion is beyond the scope of this work; nevertheless, such resonances provide a useful marker for discerning the resonance due to phase matching between the SiWG and the NGWG, which appears near 0.3\,THz when the longest NGWG passes over SiWG.  Note that the signal below 0.26\,THz is independent of the $x$-position, but close to the noise floor, because the fundamental mode of the SiWG approaches the light line (solid red line in Fig.~\ref{fig:3}(a)).

\begin{figure*}[t!]
\centering
\includegraphics[width=\textwidth]{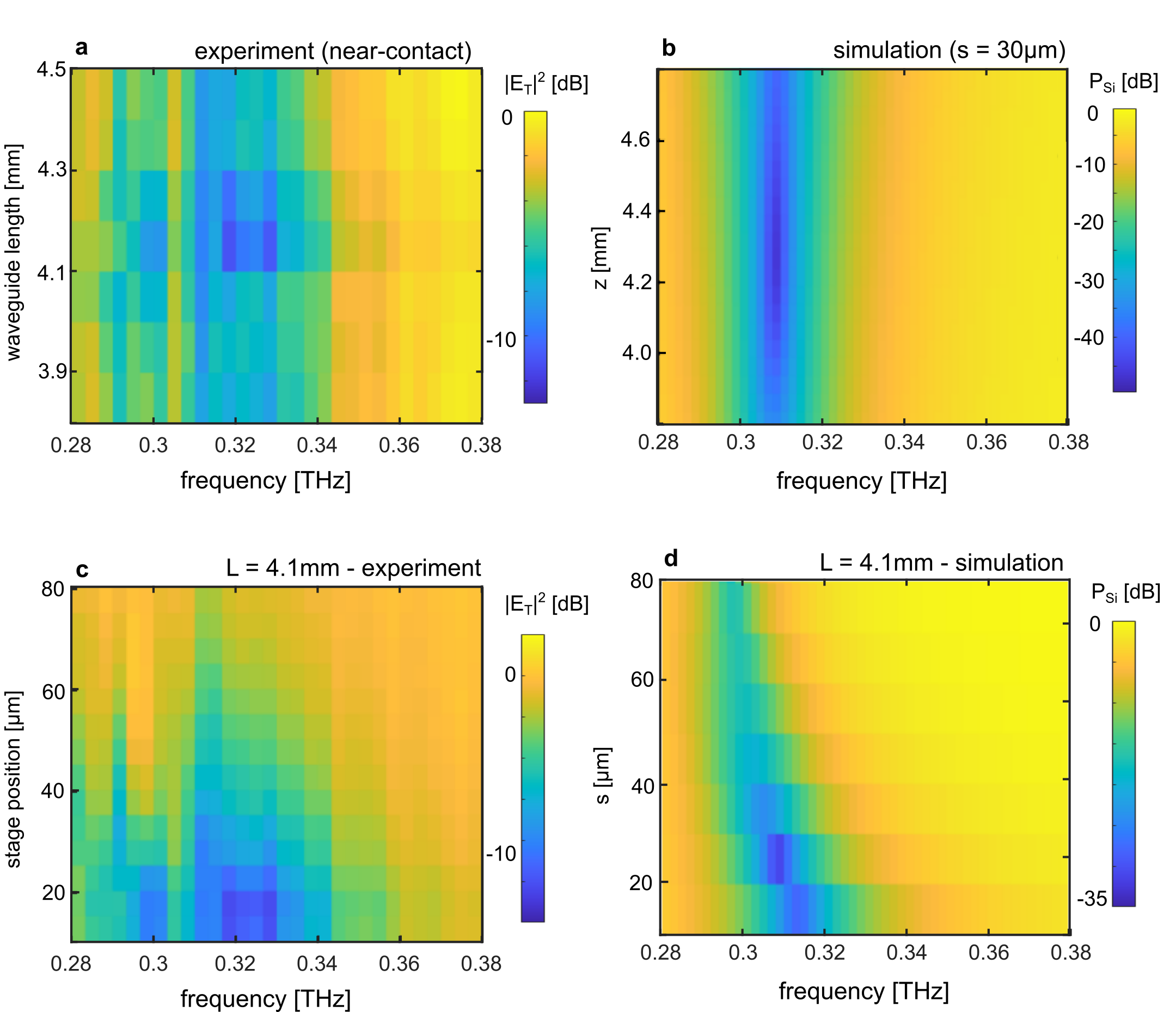}
\caption{Dependence of the nano-coupler transmission on the length $L$ of the NGWG and separation $s$ to the SiWG. (a) Measured transmitted ntensity spectrum when the SiWG and NGWG are in near contact as shown in Fig.~\ref{fig:4}(c), for different $L$. (b) Eigenmode method simulations of the power spectrum in the  SiWG when $s=30\,\mu{\rm m}$. (c) Measured transmitted intensity spectrum as a function of $s$ for $L=4.1$~mm. Note the emergence of a resonance as the waveguides are brought closer. (d) Corresponding eigenmode method simulations of the power spectrum in the dielectric waveguide as a function of $s$.}
\label{fig:5}
\end{figure*}

This first experiment indicates that resonant coupling to the NGWG mode is occurring. We now proceed with a more detailed analysis, accounting for both changes in NGWG length $L$ and separation $s$ from the edge of the SiWG. According to Fig.~\ref{fig:3}(h), in near-contact ($s=10\,\mu{\rm m}$), we expect to measure an oscillation of the power transmitted by the SiWG when changing $L$. To measure this experimentally, we repeat the transmission experiments through the SiWG, changing the length of the adjacent NGWG. Figure~\ref{fig:5}(a) shows a false colour plot of the measured transmitted intensity as a function of frequency and waveguide length $L$, where the SiWG and NGWGs are in near-contact for each measurement, and correspond to the micrographs of Fig.~\ref{fig:4}(c). We oberve that the resonance depth possesses a minimum for $L=4.1\,{\rm mm}$, and increases both for longer- and shorter- NGWG lengths, confirming the expected oscillatory behaviour (see, for example, red line in Fig.~\ref{fig:3}(h)). Figure~\ref{fig:5}(b) shows a simulation of $P_{\rm WG}$ for $s=30\,\mu{\rm m}$, i.e., the power  contained in the SiWG, showing good overall qualitative agreement with the measurements (cf.~Fig.~\ref{fig:5}(a)). By considering the $L=4.1\,{\rm mm}$ NGWG, and increasing the stage position, notice the expected gradual weakening and red-shifting of the resonance, as detailed by Fig.~\ref{fig:5}(c), as per the results of Fig.~\ref{fig:3} (see the dip in the red traces of Fig.~\ref{fig:3}(c) and \ref{fig:3}(g)). Figure~\ref{fig:5}(d) shows the corresponding simulated $P_{\rm WG}$ as a function of $s$ for $L=4.1\,{\rm mm}$, again showing good agreement with the measurements (Fig.~\ref{fig:5}(c)).

\subsection{Near-field experiments}
\begin{figure*}[t!]
\centering
\includegraphics[width=0.92\textwidth]{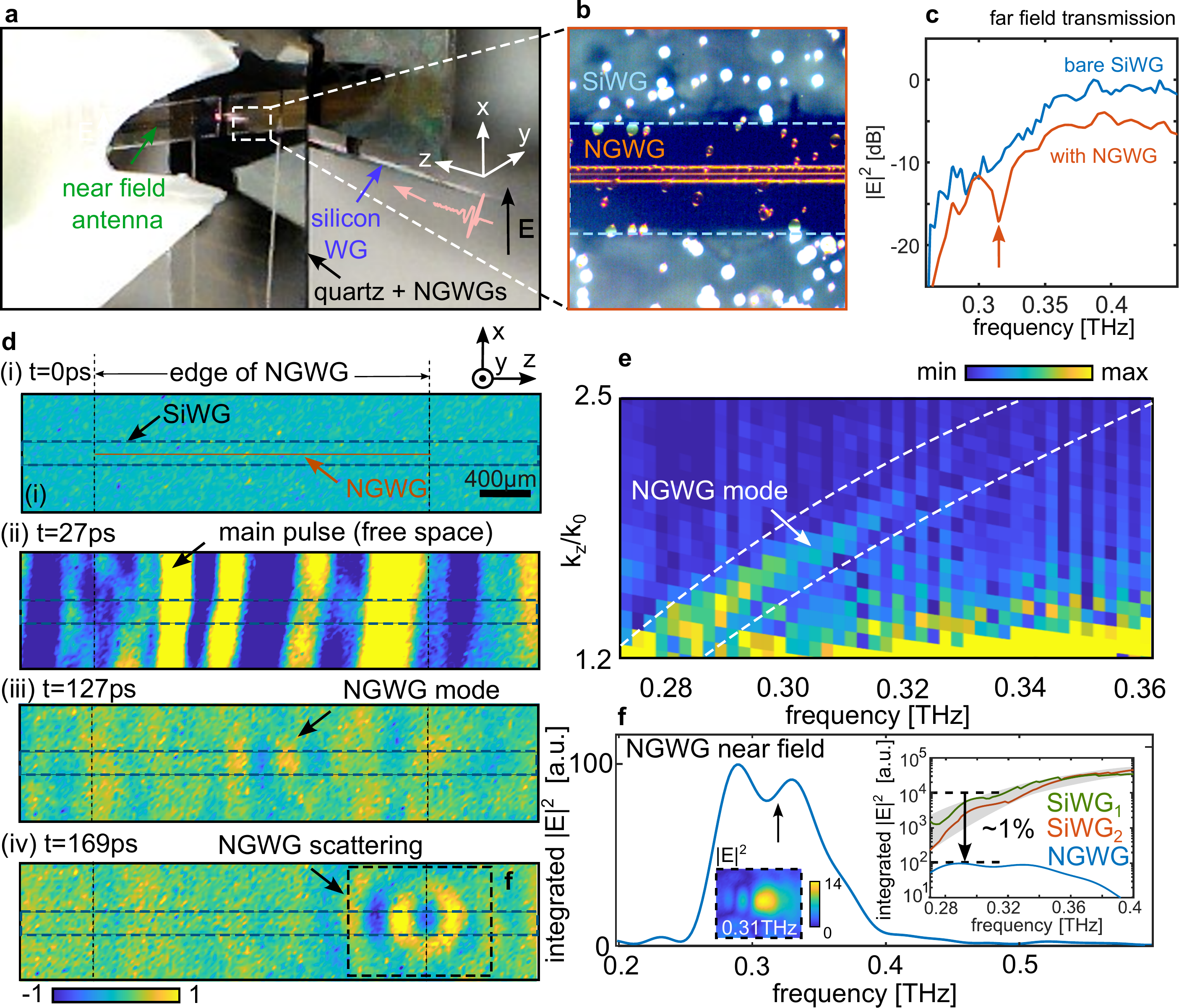}
\caption{Experimental demonstration of terahertz radiation coupling to a $\lambda/5000$ nanogap waveguide. (a) Microscope image of a terahertz near field (NF) antenna scanning the surface of a quartz substrate supporting the NGWGs, aligned with a SiWG. Note that the incoming electric field is $x$-polarized, and that the NF antenna detects $x$-polarized fields.  (b) Optical micrograph of the substrate, highlighting that the nanogap waveguide is in the center of the SiWG. (c) Far-field transmission: the red trace shows the transmitted $|E|^2$ (i.e., the far field transmission) through the silicon WG for the sample configuration shown in (a),(b). Note the resonance at 0.31\,THz, which is absent for the bare SiWG (blue trace). (d) Snapshots of the measured electric field at (i) $t=0\,$ps (no field). The horizontal lines mark the nominal SiWG boundary. The vertical dashed lines highlight the edges of the NGWG ($L=4.4\,{\rm mm}$). The orange horizontal line shows the nominal location of the NGWG, as per (b). (ii) Field snapshot at $t=27\,$ps (main pulse is from the scattered field that does not coupled to the SiWG), (iii) $t=127\,$ps (evanescent field of the NGWG mode), and (iv) $t=169\,$ps (field scattered by the end of the NGWG). The complete animation showing pulse propagation is reported in Animation 1 of the Supporting Information. (e) Measured dispersion of the detected NGWG mode using the data in (d)(iii). (f) Measured distribution of the scattered electric field intensity at the end of the NGWG using the data in (d)(iv). The scattered intensity is centered in $0.31\,$THz, confirming that the dip in (c) is due to coupling to the nanogap. Inset: comparison of the measured total intensity scattered by the nanogap (dashed box in (d)(iv)), and the estimated intensity exiting the SiWG from two independent measurements, showing percentage-level coupling efficiency. Also shown below the curve is an example of measured intensity scattered by the nanogap following a temporally gated Fourier transform, for the case of 0.31\,THz. }
\label{fig:6}
\end{figure*}

The above experiments confirm that the terahertz field at frequencies near 0.3\,THz coupled out of the SiWG if it is adjacent to the NGWG.  In a final set of experiments, we wish to confirm that such fields are indeed propagating inside the nanogap. We therefore modify the experimental setup to enable a measurement of the evanescent field, associated with the NGWG mode, via a near-field terahertz antenna~\cite{stefani2022flexible, tuniz2023subwavelength}. A photograph of the experimental setup is shown in Fig.~\ref{fig:6}(a), including the reference frame (white), direction of propagation (red) and field polarization (black). Fig.~\ref{fig:6}(b) is a top-view micrograph detail of the sample configuration, showing that the NGWG (yellow) is centered over the SiWG (dark region, bounded by the blue dashed lines). This configuration allows a far-field transmission measurement through the SiWG in first instance, followed by a near-field measurement of the electric field above the sample in second instance. The red trace in Figure~\ref{fig:6}(c) represents the field intensity spectrum transmitted through the SiWG in this configuration, and reveals, in a clearcut way, a resonance dip relative to the bare SiWG case (blue trace).  

Figure~\ref{fig:6}(d) shows representative snapshots of the measured electric field distribution in different time intervals. For clarity, panel (i) shows the SiWG boundaries as horizontal lines, whereas the vertical dashed lines highlight the edges of the NGWG ($L=4.4\,{\rm mm}$). The orange horizontal line shows the nominal location of the NGWG. The first signal to appear, at $t=27\,{\rm ps}$, is associated with the portion of the incoming pulse that does not couple to the silicon waveguide and scatters around it, as shown in Fig.~\ref{fig:6}(d)(ii). At a later time, a pulse confined to the middle of the scanning window appears, shown in Fig.~\ref{fig:6}(d)(iii) for $t=127\,{\rm ps}$, in correspondence to the location of the NGWG. We attribute this signal to the evanescent field above the quartz substrate, associated with the NGWG mode, coupled to the SiWG mode. Figure~\ref{fig:comsol} in the Supporting Information compares the relevant calculated mode profiles with- and without- the nanogap: in the absence of a nanogap, no measurable field would be present in this location. These fields move more slowly than the free space pulse, due to the larger group velocity associated with the NGWG. Once this pulse reaches the end of the NGWG, we observe the emergence of a point-dipole like emission pattern, shown in Fig.~\ref{fig:6}(d)(iv) for $t=169\,{\rm ps}$, confirming that a significant portion of the field was indeed coupled to and propagating inside the nanogap, before being scattered in the $y$ direction once the NGWG terminates. The complete and annotated animation associated with Figure~\ref{fig:6}(d) and surrounding text is shown in Animation 1 of the Supporting Information. 

Because THz-TDS measures the electric field of the terahertz pulse -- including amplitude and phase information -- we can obtain the frequency-dependent amplitude and phase in the $xz$ plane from a temporal Fourier transform. An example of the amplitude- and phase- map retrieved at 0.29\,THz is shown in Fig.~\ref{fig:spatial}. In turn, a spatial Fourier transform of the complex field $\tilde{E}(k_z,k_x)$ can be used to obtain information on the distribution of the  participating modes. For example, $|\tilde{E}(k_z,k_x)|$  at 0.29\,THz is shown in Fig.~\ref{fig:spatial}(b). Two salient spatial modes can be distinguished close to the phase-matching frequency of 0.3\,THz: the first is near $(k_z/k_0,k_x/k_0) = (1,0)$, corresponding the the main pulse of Fig.~\ref{fig:6}(d)(ii), which propagates in free space; the second is near $(k_z/k_0,k_x/k_0) = (1.6,0)$, corresponding to the later pulse propagating in the NGWG, shown in Fig.~\ref{fig:6}(d)(iii). We can confirm that this is the NGWG mode by applying a band-pass filter around the NGWG in  
$k_z/k_0 = 1.6 \pm 0.2$ and performing an inverse spatial Fourier transform. The associated field amplitude is shown in Fig.~\ref{fig:spatial}(c),  ascertaining that the field is centered in the waveguide. By binning the contribution of all $k_x$ values (i.e., by summing all rows in Fig.~\ref{fig:spatial}(b)) at each temporal frequency, we obtain a colourmap of the mode dispersion. The result is shown in Fig.~\ref{fig:6}(e): we note the excitation and detection of modes with  propagation constants above unity --  bounded by the white dashed lines -- follow the dispersion of the underlying dielectric waveguide mode (see also Fig.~\ref{fig:dielectric} in the Supporting Information). 
This overlap between the dispersions of the propagating modes measured above the nanogap and below the dielectric is an indication of  coupling  between the two waveguides.

We consider this effect in more detail, by quantifying the field scattered by the NGWG endface once the pulse reaches the end of the NGWG, corresponding to the event shown in Fig.~\ref{fig:6}(d)(iv). First, we temporally gate the data so that only the scattering event is captured (here: between $161\,{\rm ps} ~{\rm and}~190\,{\rm ps}$). Secondly, we take a temporal Fourier transform and thus obtain the field intensity $|E|^2$ at each frequency. Because the field propagates in $y$ (i.e., perpendicular to the $xz$ plane) and in free space, we can take the total field intensity to be integrated $|E|^2$  inside the $2\,{\rm mm} \times 1\,{\rm mm}$ window centered in the scattering source (black dashed rectangle in Fig.~\ref{fig:6}(d)(iv)). An example of scattered intensity distribution at 0.31\,THz is shown in the lower inset of Fig.~\ref{fig:6}. The resulting integrated $|E|^2$ as a function of frequency is shown in Fig.~\ref{fig:6}(f), indicating that its distribution peaks near 0.31\,THz, coinciding with the dielectric waveguide resonance in Fig.~\ref{fig:6}(c). This experimentally confirms that coupling from the SiWG to the NGWG has occurred.





Finally, we estimate the order of magnitude of coupling efficiency from the SiWG to the nanogap. We consider the field exiting the SiWG at the output, measured by the near-field antenna in this configuration, to obtain the estimated power propagating in $z$ and exiting the silicon waveguide. Our method uses the field distribution near the output taper of the SiWG in the $xz$ plane to estimate the field intensity propagating in $z$ in the $xy$ plane at each frequency -- see Fig.~\ref{fig:gaussian} in the Supporting Information and associated text for additional information. The resulting estimated total intensity exiting the SiWG is shown in the Fig.~\ref{fig:6}(f) inset, as red and green lines for two independent measurements, and plotted on a logarithmic scale. A comparison with the intensity scattered by the nanogap (blue line), suggests that the coupling efficiency is at least of the order of $\sim1\,\%$ with respect to the power inside the silicon waveguide. Note that this ratio considers only the measured scattered power by the nanogap which, owing to the deep subwavelength nature of the nanogap, does not emit efficiently into free space.  Preliminary calculations in COMSOL suggest that the power in the nanogap itself is at least 10 times higher than the one emitted out-of-plane, indicating that the coupling efficiency from the SiWG to the NGWG could in fact be at least 10\%. Indeed, the simulations in Fig.~\ref{fig:3}(g) indicate that the coupling efficiency could be as high as 30\%. Verifying this claim, however, would require the use of terahertz scanning near field optical microscopy inside the gap itself~\cite{guo2024terahertz}, which is a significantly more complicated experiment, and is  beyond the scope of this work.

\section{Conclusion}

In summary, we have proposed and experimentally demonstrated a directional nano-coupler capable of bridging millimeter- and nanometer- scale terahertz photonics. Our scheme implements phase-matching between a sub-mm dielectric waveguide and sub-$\mu$m metal gap waveguides, leading to resonant power transfer between them when they are placed side by side. This approach overcomes significant limitations in terms of coupling  efficiency between waveguides that differ by 3 orders of magnitude in lateral dimensions. To demonstrate the effect, we have performed a comprehensive set of complementary experiments near 0.3\,THz (i): far field experiments which measure the transmission through the THz nano-coupler; (ii) near-field experiments which measure the field associated with the nanogap waveguide. In the first set of experiments, we clearly observed the emergence of a resonant dip when the two waveguides are separated by less than 100\,$\mu$m; in the second set of experiments, we measure the propagation constant and field scattering of the mode inside the nanogap. Overall, the power scattered by the gap waveguide is $\sim1$\%, suggesting that the power in the gap itself could be more than 10\%, with simulations indicating that this could be as high as 30\%. Our results pave the way for promising developments and solutions to the problem of photonic integration at THz, with deep impact on several emerging technologies: for example, it provides a natural way for interfacing 6G terahertz circuits with photonic circuits~\cite{rajabali2023present}; it discloses new avenues for on-chip terahertz biosensing~\cite{rodrigo2015mid} at the nanoscale, e.g., to address individual proteins~\cite{yang2021near} trapped inside the nanogap; and allows new means for extreme nonlinear interaction between terahertz and optical waves on-chip~\cite{tuniz2021nanoscale} with few-layer materials possessing high nonlinearity, in the context of terahertz generation~\cite{herter2023terahertz} and detection~\cite{salamin2019compact}. Note that this approach is not limited to the specific nanogap geometry considered here: metal nanowires~\cite{yang2010theory, guo2013nanowire} and nanofilms~\cite{gacemi2013subwavelength} could couple to dielectric waveguides via the same mechanism and will likely lead to similar effects. Although metallic loss remains an outstanding issue and is the main limiting factor, these could be addressed by using lower resistivity metals or superconductors~\cite{tsiatmas2012low}.

\section{Methods}

\subsection*{Eigenmode calculations}

The propagation constants $\beta_i$, and the electric and magnetic fields ($\mathbf{e}_i$, $\mathbf{h}_i$) of the fundamental mode of each waveguide are calculated using COMSOL (Wave Optics Module, Mode Solver) -- example geometry and results are shown in Fig.~\ref{fig:3}. We use a constant  refractive index of 3.42 for silicon~\cite{dai2004terahertz}, 2.09 for quartz~\cite{davies2018temperature}, and a frequency dependent Drude model for gold~\cite{rakic1998optical}. The total electric and magnetic fields propagating in the coupler device can be written as

\begin{eqnarray}
\mathbf{E}(x,y,z) = &  a_1 {\mathbf{e}}_1(x,y) \exp(i{\beta}_1 z) 
+ a_2 {\mathbf{e}}_2(x,y) \exp(i{\beta}_2 z),  \nonumber \\
\mathbf{H}(x,y,z) = &  a_1 {\mathbf{h}}_1(x,y) \exp(i{\beta}_1 z) 
+ a_2 {\mathbf{h}}_2(x,y) \exp(i{\beta}_2 z),
\label{eq:EM2}
\end{eqnarray}
where $a_i$ is the modal complex amplitude which determines the contribution of the $i$-th eigenmode to the total field. The complex modal amplitudes are given by~\cite{tuniz2016broadband}
\begin{equation}
a_i =   \hat{z} \cdot \int \frac{1}{2} \left[{{\mathbf{e}}_i(x,y)} \times \mathbf{h}_{0}(x,y) \right] \,dxdy,
\label{eq:ai}
\end{equation}
where $\mathbf{h}_0(x,y)$ is the magnetic field of the isolated SiWG mode, corresponding to the total field at $z=0$. The total power in each waveguide is calculated by splitting the simulation space into two regions, using an artificial boundary half way between the two waveguides at $y_b=s/2$, taking the reference frame as in Fig.~\ref{fig:1}(b) with the origin at the top surface of the SiWG. We define

\begin{eqnarray}
p_{{\rm Si}}(z) &=&  \int\limits_{-\infty}^{+\infty}\int\limits_{-\infty}^{s/2} S_z(x,y)\,dxdy, \nonumber \\
p_{{\rm NG}}(z) &=& \int\limits_{-\infty}^{+\infty}\int\limits_{s/2}^{\infty} S_z(x,y)\,dxdy,
\label{eq:power_int}
\end{eqnarray}
where $p_1(z)$ and  $p_2(z)$ are the total powers (per unit length) in the SiWG and NGWG, respectively, and where $S_z$ is the $z$-component of the Poynting vector of the total field. The fraction of power in each region $p_{1}$ and $p_{2}$ respectively, as a function of propagation length, are thus defined as follows:

\begin{eqnarray}
P_{\rm Si}(z) &=& p_{\rm Si}(z)/[p_{\rm Si}(0) + p_{{\rm NG}}(0)], \nonumber \\
P_{{\rm NG}}(z) &=& p_{{\rm NG}}(z)/[p_{\rm Si}(0) + p_{{\rm NG}}(0)].
\label{eq:power}
\end{eqnarray}
so that $P_{{\rm Si}}(0) + P_{{\rm NG}}(0) = 1$. Note that, in order to resolve the field in both the NGWG and the SiWG, two integration windows are appropriately interlaced: the SiWG window is $2\,{\rm mm} \times 2\, {\rm mm}$ with a $1\,{\rm \mu m}$ pixel; the NGWG window is $4\,{\rm \mu m} \times 4\,{\rm \mu m}$ with a $2\,{\rm nm}$ pixel.

\subsection*{Sample Fabrication}

The SiWGs are fabricated by standard photo lithography and deep-reactive ion etching~\cite{bruckner2009broadband} on $250\,\mu {\rm m}$ high-resistivity float-zone silicon wafers. 
The NGWG fabrication process is as follows. A quartz substrate (thickness: $150\,\mu {\rm m}$) is piranha cleaned to remove organic contaminants; a 20\,nm layer of chromium is deposited using electron-beam evaporation. The wafer is dipped in Surpass 4000 for 60\,s, followed by DI water for 30\,s to improve the resist adhesion. ma-N 2403 photoresist is spin-coated onto the substrate at 3000 rpm, followed by baking at $90^{\circ}{\rm C}$ for 60 seconds (film thickness: 275 nm). The NGWGs are patterned via electron beam lithography (100 kV, $550~{\rm \mu C/cm}$ at 1\,nA / 100\,nA with 1\,mm write field, 2.5\,nm / 20\,nm beam step for fine and coarse structures). The exposed photoresist is developed in AZ726 for 60 seconds, followed by a DIW rinse and N$_2$ drying. A metal layer consisting of 10\,nm chromium and 100\,nm gold is deposited, followed by lift-off in dimethyl sulfoxide (DMSO) for 20 seconds at $45^{\circ}{\rm C}$. Chromium etching is carried out using Transene 1020 for 45 seconds. The etch time is calibrated using a dummy sample to achieve the desired etch depth for 30\,nm of chromium to avoid the excess delamination of the gold with Cr over-etching.

\subsection*{Far-field experiments}

We use a commercially available THz-TDS System (Menlo TERAK15) which produces THz emission from biased photoconductive antennas that are pumped by fiber-coupled near-infrared pulses (pulse width: 90\,fs; wavelength: 1560\,nm).  Polymethylpentene (TPX) lenses (Thorlabs TPX50) collimate and focus the beam towards the SiWG. The THz field emerging from the SiWG is sampled as a function of the time delay of a fiber-coupled probe pulse on another photoconductive antenna THz detector. The electric field is polarized in $x$, using the sample orientation and reference frame shown in Fig.~\ref{fig:6}. To produce the images in Fig.~\ref{fig:4} and Fig.~\ref{fig:5}, the substrate containing the NGWGs is moved via a microcontroller stage for a fixed SiWG coupling condition.

\subsection*{Near-field experiments}

The THz field on the surface of the waveguides are detected using the same technique as the far-field experiments, using another photoconductive antenna (Protemics TeraSpike TD-800-X-HR-WT-XR) which can instead scan the sample surface. Our setup measures the $x$ component of the electric field, using the reference frame of Fig.~\ref{fig:6}. A moveable, fiber-coupled near-field detector module enables the measurement of the electric field as a function of time at every point in the $xz$ plane. A second harmonic generation (SHG) module (Protemics) converts the 1560\,nm laser pulses to 780\,nm to excite the low temperature GaAs photocurrent on the antenna tip needed for terahertz detection.  Fast Fourier transforms of the temporal response at each pixel position provide the spectral information. See Animation 1 in the Supporting Information for the full measurement associated with Fig.~\ref{fig:6}(d).

\subsection*{Acknowledgments} 

This work is funded in part by the Australian Research Council Discovery Early Career Researcher Award (DE200101041). This work was performed in part at the NSW nodes of the Australian Nanofabrication Facility. We acknowledge the facilities, and the scientific and technical assistance, of the Australian Microscopy \& Microanalysis Research Facility at the Centre for Microscopy and Microanalysis, The University of Queensland. In particular, we gratefully acknowledge Gloria Qiu and Jackie He from the University of Sydney for assistance with sample fabrication. A.T. thanks Boris T. Kuhlmey for fruitful discussions.


\renewcommand{\thefigure}{S\arabic{figure}}
\setcounter{figure}{0}

\section*{Supporting Information}

\renewcommand{\thefigure}{S\arabic{figure}}
\setcounter{figure}{0}

\begin{figure*}[h!]
\centering
\includegraphics[width=0.9\textwidth]{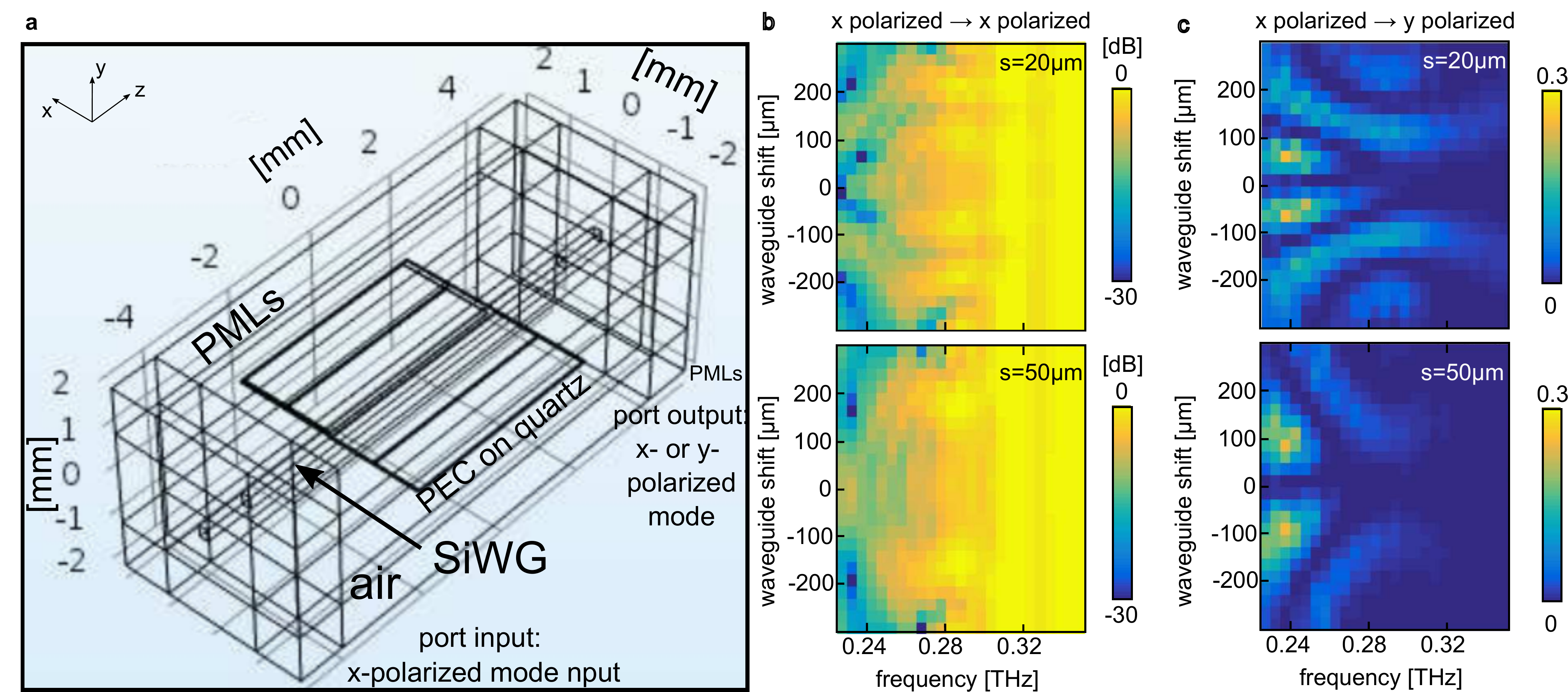}
\caption{Three dimensional COMSOL simulations, corresponding to the situation where the ``trench'' highlighted in Fig.~6(e) of the main manuscript is moved over the SiWG, and showing polarization rotation. (a) Screenshot of the simulation space, highlighting the SiWG and the quartz substrate. Here, 3D simulations are computationally feasible because the gold film is modelled via a perfect electric conductor boundary condition, where losses are ignored, which don't impose an overly onerous mesh. The port boundary conditions excite the $x$-polarized mode of the suspended SiWG at the input ($z=-4\,{\rm mm}$) one side, and detect how much power is in either the $x$ or $y$ polarized mode at the output ($z=+4\,{\rm mm}$). The geometry and material distributions aim to replicate the polarization rotation experimental conditions shown in Fig.~4. (b) Co-polarized transmitted power  (i.e.~contained in the $x$-polarization under excitation with $x$-polarization) as a function of frequency for two different values of $s$ as labelled, on a dB scale, as the waveguide is shifted over the trench. Note the good agreement with the bottom region of Fig.~4(f) of the main manuscript,  highlighted by the label ``polarization rotation''. (c) Same as (b), but for cross-polarized radiation (i.e.~considering the transmitted power contained in the $y$-polarization) as a function of frequency, on a linear scale. Note the emergence of $y$-polarized fields at lower frequencies, corresponding to polarization rotation.}
\label{fig:rotator}
\end{figure*}

\begin{figure*}[h!]
\centering
\includegraphics[width=\textwidth]{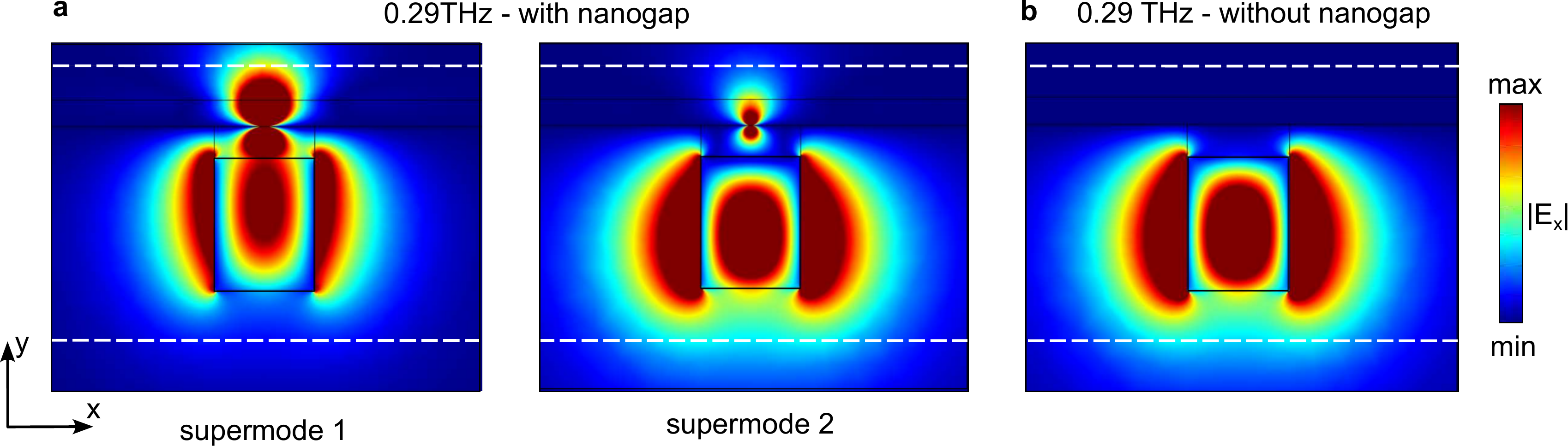}
\caption{Mode calculations showing the magnitude of the $x$ component of the electric field for relevant modes at the phase matching point, comparing the case when (a) the metallic film contains a nanogap ($w=200\,{\rm nm}$), and (b) when there is no nanogap in the film ($w=0)$. Note that material and geometric distributions are the same as in Fig.~2 of the manuscript, and that the electric field rather than the energy flux density is shown. The colourbar has been saturated to highlight how each supermode has a significantly different electric field amplitude outside the waveguide in the typical regions where the fields are measured by the near-field antenna (white dashed lines). However, note the complete absence of a measureable field in the absence of a gap in (b). We measure a field both above the NGWG (Fig.~6 of the main manuscript), and below the SiWG (see Fig.~\ref{fig:dielectric} below), corresponding to the situation in (a).}
\label{fig:comsol}
\end{figure*}

\begin{figure*}[h!]
\centering
\includegraphics[width=0.7\textwidth]{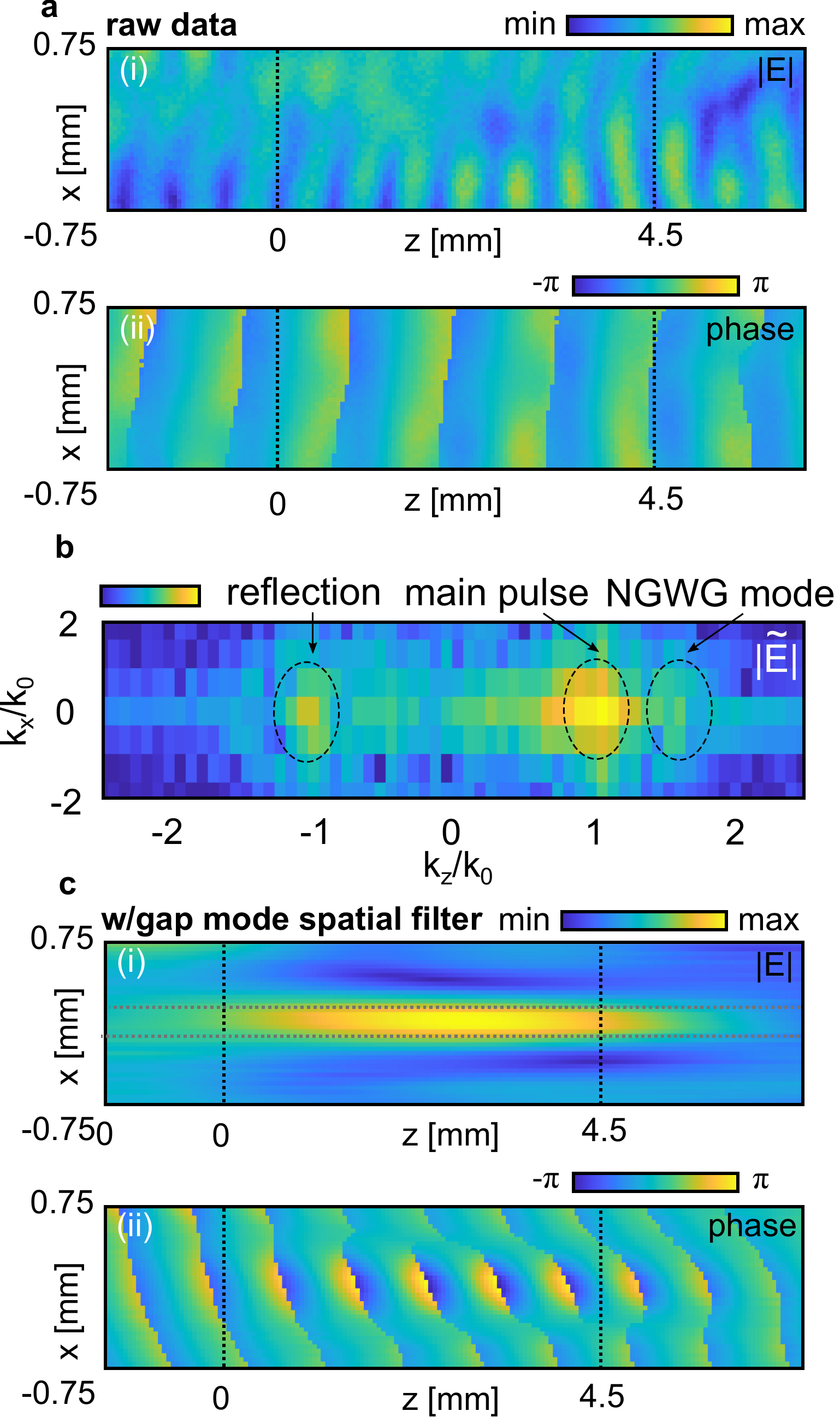}
\caption{Spatial Fourier analysis of the measured field in Fig.~6 of the main manuscript. (a) Measured distribution of the electric field magnitude (top) and phase (bottom) at 0.29\,THz. The dominant contribution originates from the scattered light in free space. (b) Magnitude of the spatial Fourier transform $|\tilde{E}|$ as a function of $k_z/k_0$ and $k_x/k_0$. The dominant contribution is from the light scattered by the waveguide in free space at $k_x/k_0\approx1$, followed by the reflected surface wave from the gold film at $k_x/k_0\approx-1$. A third, weaker contribution can be discerned at $k_x/k_0\approx1.5$. This is confirmed by binning each column, producing the dispersion in Fig.~6(e). (c) Measured field distribution obtained after a spatial band pass filter in $k_z$ in (b), bounded by the white dashed line in Fig.~6(e), followed by an inverse spatial Fourier transform. (c) The resuting field amplitude (top) and phase (bottom) reveal a NGWG mode confined within the SiWG region.}
\label{fig:spatial}
\end{figure*}

\begin{figure*}[h!]
\centering
\includegraphics[width=\textwidth]{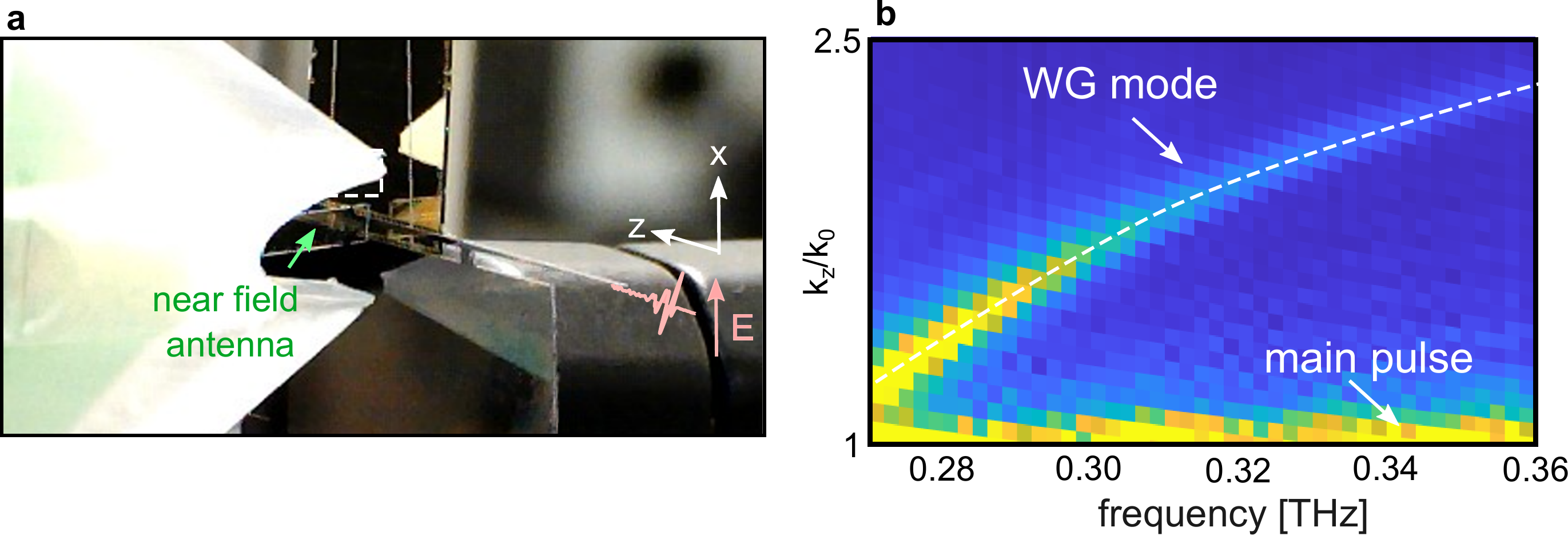}
\caption{Electric field measurements below the SiWG. (a) Microscope image of the near field antenna scanning the surface of the SiWG, where the quartz substrate supporting the NGWGs is on the opposite side. Note that the incoming electric field is $x$-polarized, and that the NF antenna detects $x$-polarized fields. (b) Measured dispersion of the detected SiWG mode using data and procedures analogous to that described in Fig.~6 of the main manuscript and surrounding text -- see also Fig.~\ref{fig:spatial} below. Note the overlap with the nanogap mode of Fig.~6 in the manuscript.}
\label{fig:dielectric}
\end{figure*}

\clearpage 

\newpage

\begin{figure*}[t!]
\centering
\includegraphics[width=0.9\textwidth]{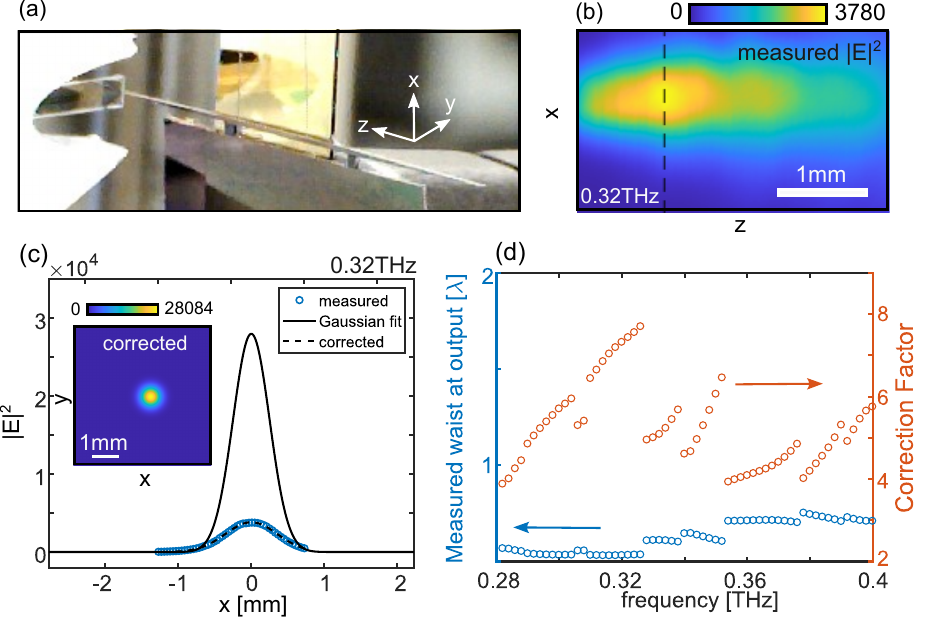}
\caption{Example of correction used in Fig. 6 of main text. (a) Microscope image of a terahertz near field antenna scanning the surface of the SiWG. (b) Measured $|E|^2$ in the region corresponding to (a). (c) Measured $|E|^2$ along the dashed line in (b) (blue circles), associated Gaussian fit (dashed line), and corrected Gaussian function. The inset shows the corresponding 2D Gaussian intensity of the beam that is estimated to exit the silicon waveguide. (d) Measured beam waist (left axis, blue circles) and associated correction factor (right axis, red circles), as described in the text.  }
\label{fig:gaussian}
\end{figure*}

\subsection*{Estimating the intensity exiting the silicon waveguide}

Because our near field measurements only scan in the $xz$ plane for a constant $y$ near the surface of each waveguide, we estimate the intensity of the electric field exiting the SiWG as follows. We first perform a near-field measurement on the surface of the SiWG at the output in the $xz$ plane under the same experimental conditions as Fig.~\ref{fig:dielectric}(a), as shown in the experimental microscope view of Fig.~\ref{fig:gaussian}(a). We therefore measure a slice of the field exiting the waveguide at constant $y$. An example field intensity at 0.32 THz, following a temporal Fourier transform inside the time interval where the field emerges from the SiWG (here: between $216\,{\rm ps}~{\rm and}~258\,{\rm ps}$), is shown in Fig.~\ref{fig:gaussian}(b). From this data, we take a 1D slice in $x$ for the value of $z$ where the intensity is maximum, highlighted by the black dashed line in Fig.~\ref{fig:gaussian}(b). The associated experimental data is shown as blue circles in Fig.~\ref{fig:gaussian}(c). We assume that this intensity is associated with a Gaussian beam, which has an intensity distribution \( I(x, y) = I_0 \exp\left(-\frac{x^2 + y^2}{2w_0^2}\right) \) centered in $(x,y) = (0,0)$. For an antenna at a distance $y=D$, \( I(x) = I_0 \exp\left(-\frac{D^2}{2w_0^2}\right) \exp\left(-\frac{x^2 }{2w_0^2}\right) \), so that 1D Gaussian fit to the measured data yields $w_0$. Knowledge of $D$ then allows us to estimate the intensity of the field exiting the waveguide propagating in $z$ -- here termed the corrected Gaussian. By way of example, the black dashed line in Fig.~\ref{fig:gaussian}(c) shows the 1D Gaussian fitted to the experimental data; the black dashed line shows the associated corrected 1D Gaussian profile; the inset shows the intensity distribution of the estimated corrected Gaussian in the $xy$ plane and propagating in $z$. We finally integrate this intensity profile to obtain the total intensity exiting the SiWG. The measured beam waist at output, in units of wavelength and as a function of frequency, is shown as blue circles (left axis) in Fig.~\ref{fig:gaussian}(d). Note that the beam waist is close to the diffraction limit. The associated correction factor, taking the estimated largest distance between the edge of the antenna and the center of the waveguide as $D=200\,\mu {\rm m}$, is shown as red circles in Fig.~6(d). This procedure produces the results shown as green lines in the inset of Fig.~6(f). Repeating the procedure in the configuration of Fig.~6(a) (not shown), using an estimated $D=700\,\mu{\rm m}$, produces comparable overall estimated total intensity exiting the SiWG, as shown in the red lines Fig.~6(f).

\bibliography{acs-main_ARXIV}

\providecommand{\latin}[1]{#1}
\makeatletter
\providecommand{\doi}
  {\begingroup\let\do\@makeother\dospecials
  \catcode`\{=1 \catcode`\}=2 \doi@aux}
\providecommand{\doi@aux}[1]{\endgroup\texttt{#1}}
\makeatother
\providecommand*\mcitethebibliography{\thebibliography}
\csname @ifundefined\endcsname{endmcitethebibliography}
  {\let\endmcitethebibliography\endthebibliography}{}
\begin{mcitethebibliography}{68}
\providecommand*\natexlab[1]{#1}
\providecommand*\mciteSetBstSublistMode[1]{}
\providecommand*\mciteSetBstMaxWidthForm[2]{}
\providecommand*\mciteBstWouldAddEndPuncttrue
  {\def\EndOfBibitem{\unskip.}}
\providecommand*\mciteBstWouldAddEndPunctfalse
  {\let\EndOfBibitem\relax}
\providecommand*\mciteSetBstMidEndSepPunct[3]{}
\providecommand*\mciteSetBstSublistLabelBeginEnd[3]{}
\providecommand*\EndOfBibitem{}
\mciteSetBstSublistMode{f}
\mciteSetBstMaxWidthForm{subitem}{(\alph{mcitesubitemcount})}
\mciteSetBstSublistLabelBeginEnd
  {\mcitemaxwidthsubitemform\space}
  {\relax}
  {\relax}

\bibitem[Markelz and Mittleman(2022)Markelz, and
  Mittleman]{markelz2022perspective}
Markelz,~A.~G.; Mittleman,~D.~M. Perspective on terahertz applications in
  bioscience and biotechnology. \emph{ACS Photonics} \textbf{2022}, \emph{9},
  1117--1126\relax
\mciteBstWouldAddEndPuncttrue
\mciteSetBstMidEndSepPunct{\mcitedefaultmidpunct}
{\mcitedefaultendpunct}{\mcitedefaultseppunct}\relax
\EndOfBibitem
\bibitem[Smolyanskaya \latin{et~al.}(2018)Smolyanskaya, Chernomyrdin, Konovko,
  Zaytsev, Ozheredov, Cherkasova, Nazarov, Guillet, Kozlov, Kistenev, Coutaz,
  Mounaix, Vaks, Son, Cheon, Wallace, Feldman, Popov, Yaroslavsky, Shkurinov,
  and Tuchin]{smolyanskaya2018terahertz}
Smolyanskaya,~O. \latin{et~al.}  Terahertz biophotonics as a tool for studies
  of dielectric and spectral properties of biological tissues and liquids.
  \emph{Progress in Quantum Electronics} \textbf{2018}, \emph{62}, 1--77\relax
\mciteBstWouldAddEndPuncttrue
\mciteSetBstMidEndSepPunct{\mcitedefaultmidpunct}
{\mcitedefaultendpunct}{\mcitedefaultseppunct}\relax
\EndOfBibitem
\bibitem[Seo and Kim(2022)Seo, and Kim]{Seo2022}
Seo,~C.; Kim,~T.-T. Terahertz near-field spectroscopy for various applications.
  \emph{Journal of the Korean Physical Society} \textbf{2022}, \emph{81},
  549--561\relax
\mciteBstWouldAddEndPuncttrue
\mciteSetBstMidEndSepPunct{\mcitedefaultmidpunct}
{\mcitedefaultendpunct}{\mcitedefaultseppunct}\relax
\EndOfBibitem
\bibitem[Siegel(2007)]{siegel2007thz}
Siegel,~P.~H. THz instruments for space. \emph{IEEE Transactions on Antennas
  and Propagation} \textbf{2007}, \emph{55}, 2957--2965\relax
\mciteBstWouldAddEndPuncttrue
\mciteSetBstMidEndSepPunct{\mcitedefaultmidpunct}
{\mcitedefaultendpunct}{\mcitedefaultseppunct}\relax
\EndOfBibitem
\bibitem[Kleine-Ostmann and Nagatsuma(2011)Kleine-Ostmann, and
  Nagatsuma]{kleine2011review}
Kleine-Ostmann,~T.; Nagatsuma,~T. A review on terahertz communications
  research. \emph{Journal of Infrared, Millimeter, and Terahertz Waves}
  \textbf{2011}, \emph{32}, 143--171\relax
\mciteBstWouldAddEndPuncttrue
\mciteSetBstMidEndSepPunct{\mcitedefaultmidpunct}
{\mcitedefaultendpunct}{\mcitedefaultseppunct}\relax
\EndOfBibitem
\bibitem[Sizov and Rogalski(2010)Sizov, and Rogalski]{sizov2010thz}
Sizov,~F.; Rogalski,~A. THz detectors. \emph{Progress in quantum electronics}
  \textbf{2010}, \emph{34}, 278--347\relax
\mciteBstWouldAddEndPuncttrue
\mciteSetBstMidEndSepPunct{\mcitedefaultmidpunct}
{\mcitedefaultendpunct}{\mcitedefaultseppunct}\relax
\EndOfBibitem
\bibitem[Pang \latin{et~al.}(2022)Pang, Ozolins, Jia, Zhang, Schatz, Udalcovs,
  Bobrovs, Hu, Morioka, Sun, \latin{et~al.} others]{pang2022bridging}
Pang,~X.; Ozolins,~O.; Jia,~S.; Zhang,~L.; Schatz,~R.; Udalcovs,~A.;
  Bobrovs,~V.; Hu,~H.; Morioka,~T.; Sun,~Y.-T.; others Bridging the terahertz
  gap: Photonics-assisted free-space communications from the submillimeter-wave
  to the mid-infrared. \emph{Journal of Lightwave Technology} \textbf{2022},
  \emph{40}, 3149--3162\relax
\mciteBstWouldAddEndPuncttrue
\mciteSetBstMidEndSepPunct{\mcitedefaultmidpunct}
{\mcitedefaultendpunct}{\mcitedefaultseppunct}\relax
\EndOfBibitem
\bibitem[Headland \latin{et~al.}(2020)Headland, Withayachumnankul, Yu, Fujita,
  and Nagatsuma]{headland2020unclad}
Headland,~D.; Withayachumnankul,~W.; Yu,~X.; Fujita,~M.; Nagatsuma,~T. Unclad
  microphotonics for terahertz waveguides and systems. \emph{Journal of
  Lightwave Technology} \textbf{2020}, \emph{38}, 6853--6862\relax
\mciteBstWouldAddEndPuncttrue
\mciteSetBstMidEndSepPunct{\mcitedefaultmidpunct}
{\mcitedefaultendpunct}{\mcitedefaultseppunct}\relax
\EndOfBibitem
\bibitem[Xu and Skorobogatiy(2022)Xu, and Skorobogatiy]{xu2022wired}
Xu,~G.; Skorobogatiy,~M. Wired {THz} Communications. \emph{Journal of Infrared,
  Millimeter, and Terahertz Waves} \textbf{2022}, \emph{43}, 728--778\relax
\mciteBstWouldAddEndPuncttrue
\mciteSetBstMidEndSepPunct{\mcitedefaultmidpunct}
{\mcitedefaultendpunct}{\mcitedefaultseppunct}\relax
\EndOfBibitem
\bibitem[Lawler \latin{et~al.}(2020)Lawler, Ho, Evans, Wallace, and
  Iyer]{lawler2020convergence}
Lawler,~N.~B.; Ho,~D.; Evans,~C.~W.; Wallace,~V.~P.; Iyer,~K.~S. Convergence of
  terahertz radiation and nanotechnology. \emph{Journal of Materials Chemistry
  C} \textbf{2020}, \emph{8}, 10942--10955\relax
\mciteBstWouldAddEndPuncttrue
\mciteSetBstMidEndSepPunct{\mcitedefaultmidpunct}
{\mcitedefaultendpunct}{\mcitedefaultseppunct}\relax
\EndOfBibitem
\bibitem[Leitenstorfer \latin{et~al.}(2023)Leitenstorfer, Moskalenko,
  Kampfrath, Kono, Castro-Camus, Peng, Qureshi, Turchinovich, Tanaka, Markelz,
  \latin{et~al.} others]{leitenstorfer20232023}
Leitenstorfer,~A.; Moskalenko,~A.~S.; Kampfrath,~T.; Kono,~J.;
  Castro-Camus,~E.; Peng,~K.; Qureshi,~N.; Turchinovich,~D.; Tanaka,~K.;
  Markelz,~A.~G.; others The 2023 terahertz science and technology roadmap.
  \emph{Journal of Physics D: Applied Physics} \textbf{2023}, \emph{56},
  223001\relax
\mciteBstWouldAddEndPuncttrue
\mciteSetBstMidEndSepPunct{\mcitedefaultmidpunct}
{\mcitedefaultendpunct}{\mcitedefaultseppunct}\relax
\EndOfBibitem
\bibitem[Rosei(2004)]{rosei2004nanostructured}
Rosei,~F. Nanostructured surfaces: challenges and frontiers in nanotechnology.
  \emph{Journal of Physics: Condensed Matter} \textbf{2004}, \emph{16},
  S1373\relax
\mciteBstWouldAddEndPuncttrue
\mciteSetBstMidEndSepPunct{\mcitedefaultmidpunct}
{\mcitedefaultendpunct}{\mcitedefaultseppunct}\relax
\EndOfBibitem
\bibitem[Chen \latin{et~al.}(2003)Chen, Kersting, and Cho]{chen2003terahertz}
Chen,~H.-T.; Kersting,~R.; Cho,~G.~C. Terahertz imaging with nanometer
  resolution. \emph{Applied Physics Letters} \textbf{2003}, \emph{83},
  3009--3011\relax
\mciteBstWouldAddEndPuncttrue
\mciteSetBstMidEndSepPunct{\mcitedefaultmidpunct}
{\mcitedefaultendpunct}{\mcitedefaultseppunct}\relax
\EndOfBibitem
\bibitem[Wittmann \latin{et~al.}(2023)Wittmann, Pindl, Sawallich, Nagel,
  Michalski, Pandey, Esteki, Kataria, and Lemme]{wittmann2023assessment}
Wittmann,~S.; Pindl,~S.; Sawallich,~S.; Nagel,~M.; Michalski,~A.; Pandey,~H.;
  Esteki,~A.; Kataria,~S.; Lemme,~M.~C. Assessment of wafer-level transfer
  techniques of graphene with respect to semiconductor industry requirements.
  \emph{Advanced Materials Technologies} \textbf{2023}, \emph{8}, 2201587\relax
\mciteBstWouldAddEndPuncttrue
\mciteSetBstMidEndSepPunct{\mcitedefaultmidpunct}
{\mcitedefaultendpunct}{\mcitedefaultseppunct}\relax
\EndOfBibitem
\bibitem[Tuniz and Kuhlmey(2023)Tuniz, and Kuhlmey]{tuniz2023subwavelength}
Tuniz,~A.; Kuhlmey,~B.~T. Subwavelength terahertz imaging via virtual
  superlensing in the radiating near field. \emph{Nature Communications}
  \textbf{2023}, \emph{14}, 6393\relax
\mciteBstWouldAddEndPuncttrue
\mciteSetBstMidEndSepPunct{\mcitedefaultmidpunct}
{\mcitedefaultendpunct}{\mcitedefaultseppunct}\relax
\EndOfBibitem
\bibitem[Guo \latin{et~al.}(2024)Guo, Bertling, Donose, Br{\"u}nig, Cernescu,
  Govyadinov, and Raki{\'c}]{guo2024terahertz}
Guo,~X.; Bertling,~K.; Donose,~B.~C.; Br{\"u}nig,~M.; Cernescu,~A.;
  Govyadinov,~A.~A.; Raki{\'c},~A.~D. Terahertz nanoscopy: Advances,
  challenges, and the road ahead. \emph{Applied Physics Reviews} \textbf{2024},
  \emph{11}\relax
\mciteBstWouldAddEndPuncttrue
\mciteSetBstMidEndSepPunct{\mcitedefaultmidpunct}
{\mcitedefaultendpunct}{\mcitedefaultseppunct}\relax
\EndOfBibitem
\bibitem[Cocker \latin{et~al.}(2021)Cocker, Jelic, Hillenbrand, and
  Hegmann]{cocker2021nanoscale}
Cocker,~T.; Jelic,~V.; Hillenbrand,~R.; Hegmann,~F. Nanoscale terahertz
  scanning probe microscopy. \emph{Nature Photonics} \textbf{2021}, \emph{15},
  558--569\relax
\mciteBstWouldAddEndPuncttrue
\mciteSetBstMidEndSepPunct{\mcitedefaultmidpunct}
{\mcitedefaultendpunct}{\mcitedefaultseppunct}\relax
\EndOfBibitem
\bibitem[Kaltenecker \latin{et~al.}(2021)Kaltenecker, G{\"o}lz, Bau, and
  Keilmann]{kaltenecker2021infrared}
Kaltenecker,~K.~J.; G{\"o}lz,~T.; Bau,~E.; Keilmann,~F. Infrared-spectroscopic,
  dynamic near-field microscopy of living cells and nanoparticles in water.
  \emph{Scientific Reports} \textbf{2021}, \emph{11}, 21860\relax
\mciteBstWouldAddEndPuncttrue
\mciteSetBstMidEndSepPunct{\mcitedefaultmidpunct}
{\mcitedefaultendpunct}{\mcitedefaultseppunct}\relax
\EndOfBibitem
\bibitem[Park \latin{et~al.}(2012)Park, Jin, Yi, Ye, Ahn, and
  Jeong]{park2012enhancement}
Park,~S.-G.; Jin,~K.~H.; Yi,~M.; Ye,~J.~C.; Ahn,~J.; Jeong,~K.-H. Enhancement
  of terahertz pulse emission by optical nanoantenna. \emph{ACS Nano}
  \textbf{2012}, \emph{6}, 2026--2031\relax
\mciteBstWouldAddEndPuncttrue
\mciteSetBstMidEndSepPunct{\mcitedefaultmidpunct}
{\mcitedefaultendpunct}{\mcitedefaultseppunct}\relax
\EndOfBibitem
\bibitem[Peters \latin{et~al.}(2024)Peters, Rocco, Olivieri, Arregui~Leon,
  Cecconi, Carletti, Gigli, Della~Valle, Cutrona, Totero~Gongora, Leo,
  Pasquazi, De~Angelis, and Peccianti]{PetersAOM2024}
Peters,~L.; Rocco,~D.; Olivieri,~L.; Arregui~Leon,~U.; Cecconi,~V.;
  Carletti,~L.; Gigli,~C.; Della~Valle,~G.; Cutrona,~A.; Totero~Gongora,~J.~S.;
  Leo,~G.; Pasquazi,~A.; De~Angelis,~C.; Peccianti,~M. Resonant Fully
  Dielectric Metasurfaces for Ultrafast Terahertz Pulse Generation.
  \emph{Advanced Optical Materials} \textbf{2024}, \emph{12}, 2303148\relax
\mciteBstWouldAddEndPuncttrue
\mciteSetBstMidEndSepPunct{\mcitedefaultmidpunct}
{\mcitedefaultendpunct}{\mcitedefaultseppunct}\relax
\EndOfBibitem
\bibitem[Vitiello \latin{et~al.}(2012)Vitiello, Coquillat, Viti, Ercolani,
  Teppe, Pitanti, Beltram, Sorba, Knap, and Tredicucci]{vitiello2012room}
Vitiello,~M.~S.; Coquillat,~D.; Viti,~L.; Ercolani,~D.; Teppe,~F.; Pitanti,~A.;
  Beltram,~F.; Sorba,~L.; Knap,~W.; Tredicucci,~A. Room-temperature terahertz
  detectors based on semiconductor nanowire field-effect transistors.
  \emph{Nano Letters} \textbf{2012}, \emph{12}, 96--101\relax
\mciteBstWouldAddEndPuncttrue
\mciteSetBstMidEndSepPunct{\mcitedefaultmidpunct}
{\mcitedefaultendpunct}{\mcitedefaultseppunct}\relax
\EndOfBibitem
\bibitem[Degl’Innocenti \latin{et~al.}(2018)Degl’Innocenti, Kindness,
  Beere, and Ritchie]{degl2018all}
Degl’Innocenti,~R.; Kindness,~S.~J.; Beere,~H.~E.; Ritchie,~D.~A.
  All-integrated terahertz modulators. \emph{Nanophotonics} \textbf{2018},
  \emph{7}, 127--144\relax
\mciteBstWouldAddEndPuncttrue
\mciteSetBstMidEndSepPunct{\mcitedefaultmidpunct}
{\mcitedefaultendpunct}{\mcitedefaultseppunct}\relax
\EndOfBibitem
\bibitem[Headland \latin{et~al.}(2023)Headland, Fujita, Carpintero, Nagatsuma,
  and Withayachumnankul]{headland2023terahertz}
Headland,~D.; Fujita,~M.; Carpintero,~G.; Nagatsuma,~T.; Withayachumnankul,~W.
  Terahertz integration platforms using substrateless all-silicon
  microstructures. \emph{APL Photonics} \textbf{2023}, \emph{8}\relax
\mciteBstWouldAddEndPuncttrue
\mciteSetBstMidEndSepPunct{\mcitedefaultmidpunct}
{\mcitedefaultendpunct}{\mcitedefaultseppunct}\relax
\EndOfBibitem
\bibitem[Han \latin{et~al.}(2015)Han, Zhang, and Bozhevolnyi]{han2015spoof}
Han,~Z.; Zhang,~Y.; Bozhevolnyi,~S.~I. Spoof surface plasmon-based stripe
  antennas with extreme field enhancement in the terahertz regime. \emph{Optics
  Letters} \textbf{2015}, \emph{40}, 2533--2536\relax
\mciteBstWouldAddEndPuncttrue
\mciteSetBstMidEndSepPunct{\mcitedefaultmidpunct}
{\mcitedefaultendpunct}{\mcitedefaultseppunct}\relax
\EndOfBibitem
\bibitem[Tsiatmas \latin{et~al.}(2012)Tsiatmas, Fedotov, de~Abajo, and
  Zheludev]{tsiatmas2012low}
Tsiatmas,~A.; Fedotov,~V.~A.; de~Abajo,~F. J.~G.; Zheludev,~N.~I. Low-loss
  terahertz superconducting plasmonics. \emph{New Journal of Physics}
  \textbf{2012}, \emph{14}, 115006\relax
\mciteBstWouldAddEndPuncttrue
\mciteSetBstMidEndSepPunct{\mcitedefaultmidpunct}
{\mcitedefaultendpunct}{\mcitedefaultseppunct}\relax
\EndOfBibitem
\bibitem[Yang \latin{et~al.}(2010)Yang, Cao, and Zhou]{yang2010theory}
Yang,~J.; Cao,~Q.; Zhou,~C. Theory for terahertz plasmons of metallic nanowires
  with sub-skin-depth diameters. \emph{Optics Express} \textbf{2010},
  \emph{18}, 18550--18557\relax
\mciteBstWouldAddEndPuncttrue
\mciteSetBstMidEndSepPunct{\mcitedefaultmidpunct}
{\mcitedefaultendpunct}{\mcitedefaultseppunct}\relax
\EndOfBibitem
\bibitem[Salamin \latin{et~al.}(2019)Salamin, Benea-Chelmus, Fedoryshyn, Heni,
  Elder, Dalton, Faist, and Leuthold]{salamin2019compact}
Salamin,~Y.; Benea-Chelmus,~I.-C.; Fedoryshyn,~Y.; Heni,~W.; Elder,~D.~L.;
  Dalton,~L.~R.; Faist,~J.; Leuthold,~J. Compact and ultra-efficient broadband
  plasmonic terahertz field detector. \emph{Nature Communications}
  \textbf{2019}, \emph{10}, 5550\relax
\mciteBstWouldAddEndPuncttrue
\mciteSetBstMidEndSepPunct{\mcitedefaultmidpunct}
{\mcitedefaultendpunct}{\mcitedefaultseppunct}\relax
\EndOfBibitem
\bibitem[Sengupta \latin{et~al.}(2018)Sengupta, Nagatsuma, and
  Mittleman]{sengupta2018terahertz}
Sengupta,~K.; Nagatsuma,~T.; Mittleman,~D.~M. Terahertz integrated electronic
  and hybrid electronic--photonic systems. \emph{Nature Electronics}
  \textbf{2018}, \emph{1}, 622--635\relax
\mciteBstWouldAddEndPuncttrue
\mciteSetBstMidEndSepPunct{\mcitedefaultmidpunct}
{\mcitedefaultendpunct}{\mcitedefaultseppunct}\relax
\EndOfBibitem
\bibitem[Rajabali and Benea-Chelmus(2023)Rajabali, and
  Benea-Chelmus]{rajabali2023present}
Rajabali,~S.; Benea-Chelmus,~I.-C. Present and future of terahertz integrated
  photonic devices. \emph{{APL} Photonics} \textbf{2023}, \emph{8}\relax
\mciteBstWouldAddEndPuncttrue
\mciteSetBstMidEndSepPunct{\mcitedefaultmidpunct}
{\mcitedefaultendpunct}{\mcitedefaultseppunct}\relax
\EndOfBibitem
\bibitem[Park \latin{et~al.}(2015)Park, Chen, Nguyen, Peraire, and
  Oh]{park2015nanogap}
Park,~H.-R.; Chen,~X.; Nguyen,~N.-C.; Peraire,~J.; Oh,~S.-H. Nanogap-enhanced
  terahertz sensing of 1 nm thick ($\lambda$/106) dielectric films. \emph{ACS
  Photonics} \textbf{2015}, \emph{2}, 417--424\relax
\mciteBstWouldAddEndPuncttrue
\mciteSetBstMidEndSepPunct{\mcitedefaultmidpunct}
{\mcitedefaultendpunct}{\mcitedefaultseppunct}\relax
\EndOfBibitem
\bibitem[Park \latin{et~al.}(2017)Park, Cha, Shin, and Ahn]{park2017sensing}
Park,~S.; Cha,~S.; Shin,~G.; Ahn,~Y. Sensing viruses using terahertz nano-gap
  metamaterials. \emph{Biomedical Optics Express} \textbf{2017}, \emph{8},
  3551--3558\relax
\mciteBstWouldAddEndPuncttrue
\mciteSetBstMidEndSepPunct{\mcitedefaultmidpunct}
{\mcitedefaultendpunct}{\mcitedefaultseppunct}\relax
\EndOfBibitem
\bibitem[Kim \latin{et~al.}(2018)Kim, In, Lee, Rhie, Jeong, Kim, and
  Park]{kim2018colossal}
Kim,~N.; In,~S.; Lee,~D.; Rhie,~J.; Jeong,~J.; Kim,~D.-S.; Park,~N. Colossal
  terahertz field enhancement using split-ring resonators with a sub-10 nm gap.
  \emph{{ACS} Photonics} \textbf{2018}, \emph{5}, 278--283\relax
\mciteBstWouldAddEndPuncttrue
\mciteSetBstMidEndSepPunct{\mcitedefaultmidpunct}
{\mcitedefaultendpunct}{\mcitedefaultseppunct}\relax
\EndOfBibitem
\bibitem[Lee \latin{et~al.}(2015)Lee, Kang, Lee, Kim, Kim, Hun~Kim, Lee, Son,
  Park, and Seo]{lee2015highly}
Lee,~D.-K.; Kang,~J.-H.; Lee,~J.-S.; Kim,~H.-S.; Kim,~C.; Hun~Kim,~J.; Lee,~T.;
  Son,~J.-H.; Park,~Q.-H.; Seo,~M. Highly sensitive and selective sugar
  detection by terahertz nano-antennas. \emph{Scientific Reports}
  \textbf{2015}, \emph{5}, 15459\relax
\mciteBstWouldAddEndPuncttrue
\mciteSetBstMidEndSepPunct{\mcitedefaultmidpunct}
{\mcitedefaultendpunct}{\mcitedefaultseppunct}\relax
\EndOfBibitem
\bibitem[Bahk \latin{et~al.}(2019)Bahk, Kim, and Park]{bahk2019large}
Bahk,~Y.-M.; Kim,~D.-S.; Park,~H.-R. Large-Area Metal Gaps and Their Optical
  Applications. \emph{Advanced Optical Materials} \textbf{2019}, \emph{7},
  1800426\relax
\mciteBstWouldAddEndPuncttrue
\mciteSetBstMidEndSepPunct{\mcitedefaultmidpunct}
{\mcitedefaultendpunct}{\mcitedefaultseppunct}\relax
\EndOfBibitem
\bibitem[Stegeman \latin{et~al.}(1983)Stegeman, Wallis, and
  Maradudin]{stegeman1983excitation}
Stegeman,~G.; Wallis,~R.; Maradudin,~A. Excitation of surface polaritons by
  end-fire coupling. \emph{Optics Letters} \textbf{1983}, \emph{8},
  386--388\relax
\mciteBstWouldAddEndPuncttrue
\mciteSetBstMidEndSepPunct{\mcitedefaultmidpunct}
{\mcitedefaultendpunct}{\mcitedefaultseppunct}\relax
\EndOfBibitem
\bibitem[Taras \latin{et~al.}(2021)Taras, Tuniz, Bajwa, Ng, Dawes, Poulton, and
  {d}e Sterke]{taras2021shortcuts}
Taras,~A.~K.; Tuniz,~A.; Bajwa,~M.~A.; Ng,~V.; Dawes,~J.~M.; Poulton,~C.~G.;
  {d}e Sterke,~C.~M. Shortcuts to adiabaticity in waveguide couplers--theory
  and implementation. \emph{Advances in Physics: X} \textbf{2021}, \emph{6},
  1894978\relax
\mciteBstWouldAddEndPuncttrue
\mciteSetBstMidEndSepPunct{\mcitedefaultmidpunct}
{\mcitedefaultendpunct}{\mcitedefaultseppunct}\relax
\EndOfBibitem
\bibitem[Tuniz \latin{et~al.}(2024)Tuniz, Song, Della~Valle, and
  de~Sterke]{tuniz2024coupled}
Tuniz,~A.; Song,~A.~Y.; Della~Valle,~G.; de~Sterke,~C.~M. Coupled mode theory
  for plasmonic couplers. \emph{Applied Physics Reviews} \textbf{2024},
  \emph{11}\relax
\mciteBstWouldAddEndPuncttrue
\mciteSetBstMidEndSepPunct{\mcitedefaultmidpunct}
{\mcitedefaultendpunct}{\mcitedefaultseppunct}\relax
\EndOfBibitem
\bibitem[Yu \latin{et~al.}(2019)Yu, Kim, Fujita, and
  Nagatsuma]{yu2019efficient}
Yu,~X.; Kim,~J.-Y.; Fujita,~M.; Nagatsuma,~T. Efficient mode converter to
  deep-subwavelength region with photonic-crystal waveguide platform for
  terahertz applications. \emph{Optics Express} \textbf{2019}, \emph{27},
  28707--28721\relax
\mciteBstWouldAddEndPuncttrue
\mciteSetBstMidEndSepPunct{\mcitedefaultmidpunct}
{\mcitedefaultendpunct}{\mcitedefaultseppunct}\relax
\EndOfBibitem
\bibitem[Delacour \latin{et~al.}(2010)Delacour, Blaize, Grosse, Fedeli,
  Bruyant, Salas-Montiel, Lerondel, and Chelnokov]{delacour2010efficient}
Delacour,~C.; Blaize,~S.; Grosse,~P.; Fedeli,~J.~M.; Bruyant,~A.;
  Salas-Montiel,~R.; Lerondel,~G.; Chelnokov,~A. Efficient directional coupling
  between silicon and copper plasmonic nanoslot waveguides: toward metal-
  oxide- silicon nanophotonics. \emph{Nano Letters} \textbf{2010}, \emph{10},
  2922--2926\relax
\mciteBstWouldAddEndPuncttrue
\mciteSetBstMidEndSepPunct{\mcitedefaultmidpunct}
{\mcitedefaultendpunct}{\mcitedefaultseppunct}\relax
\EndOfBibitem
\bibitem[Nielsen \latin{et~al.}(2017)Nielsen, Shi, Dichtl, Maier, and
  Oulton]{nielsen2017giant}
Nielsen,~M.~P.; Shi,~X.; Dichtl,~P.; Maier,~S.~A.; Oulton,~R.~F. Giant
  nonlinear response at a plasmonic nanofocus drives efficient four-wave
  mixing. \emph{Science} \textbf{2017}, \emph{358}, 1179--1181\relax
\mciteBstWouldAddEndPuncttrue
\mciteSetBstMidEndSepPunct{\mcitedefaultmidpunct}
{\mcitedefaultendpunct}{\mcitedefaultseppunct}\relax
\EndOfBibitem
\bibitem[Novotny and Hecht(2012)Novotny, and Hecht]{novotny2012principles}
Novotny,~L.; Hecht,~B. \emph{Principles of nano-optics}; Cambridge University
  Press: Cambridge, 2012\relax
\mciteBstWouldAddEndPuncttrue
\mciteSetBstMidEndSepPunct{\mcitedefaultmidpunct}
{\mcitedefaultendpunct}{\mcitedefaultseppunct}\relax
\EndOfBibitem
\bibitem[Seo \latin{et~al.}(2009)Seo, Park, Koo, Park, Kang, Suwal, Choi,
  Planken, Park, Park, \latin{et~al.} others]{seo2009terahertz}
Seo,~M.; Park,~H.; Koo,~S.; Park,~D.; Kang,~J.; Suwal,~O.; Choi,~S.;
  Planken,~P.; Park,~G.; Park,~N.; others Terahertz field enhancement by a
  metallic nano slit operating beyond the skin-depth limit. \emph{Nature
  Photonics} \textbf{2009}, \emph{3}, 152--156\relax
\mciteBstWouldAddEndPuncttrue
\mciteSetBstMidEndSepPunct{\mcitedefaultmidpunct}
{\mcitedefaultendpunct}{\mcitedefaultseppunct}\relax
\EndOfBibitem
\bibitem[He(2009)]{he2009investigation}
He,~X.-Y. Investigation of terahertz Sommerfeld wave propagation along conical
  metal wire. \emph{JOSA B} \textbf{2009}, \emph{26}, A23--A28\relax
\mciteBstWouldAddEndPuncttrue
\mciteSetBstMidEndSepPunct{\mcitedefaultmidpunct}
{\mcitedefaultendpunct}{\mcitedefaultseppunct}\relax
\EndOfBibitem
\bibitem[Dai \latin{et~al.}(2004)Dai, Zhang, Zhang, and
  Grischkowsky]{dai2004terahertz}
Dai,~J.; Zhang,~J.; Zhang,~W.; Grischkowsky,~D. Terahertz time-domain
  spectroscopy characterization of the far-infrared absorption and index of
  refraction of high-resistivity, float-zone silicon. \emph{Journal of the
  Optical Society of America B} \textbf{2004}, \emph{21}, 1379--1386\relax
\mciteBstWouldAddEndPuncttrue
\mciteSetBstMidEndSepPunct{\mcitedefaultmidpunct}
{\mcitedefaultendpunct}{\mcitedefaultseppunct}\relax
\EndOfBibitem
\bibitem[Davies \latin{et~al.}(2018)Davies, Patel, Xia, Herz, and
  Johnston]{davies2018temperature}
Davies,~C.~L.; Patel,~J.~B.; Xia,~C.~Q.; Herz,~L.~M.; Johnston,~M.~B.
  Temperature-dependent refractive index of quartz at terahertz frequencies.
  \emph{Journal of Infrared, Millimeter, and Terahertz Waves} \textbf{2018},
  \emph{39}, 1236--1248\relax
\mciteBstWouldAddEndPuncttrue
\mciteSetBstMidEndSepPunct{\mcitedefaultmidpunct}
{\mcitedefaultendpunct}{\mcitedefaultseppunct}\relax
\EndOfBibitem
\bibitem[Raki{\'c} \latin{et~al.}(1998)Raki{\'c}, Djuri{\v{s}}i{\'c}, Elazar,
  and Majewski]{rakic1998optical}
Raki{\'c},~A.~D.; Djuri{\v{s}}i{\'c},~A.~B.; Elazar,~J.~M.; Majewski,~M.~L.
  Optical properties of metallic films for vertical-cavity optoelectronic
  devices. \emph{Applied Optics} \textbf{1998}, \emph{37}, 5271--5283\relax
\mciteBstWouldAddEndPuncttrue
\mciteSetBstMidEndSepPunct{\mcitedefaultmidpunct}
{\mcitedefaultendpunct}{\mcitedefaultseppunct}\relax
\EndOfBibitem
\bibitem[Huang(1994)]{huang1994coupled}
Huang,~W.-P. Coupled-mode theory for optical waveguides: an overview. \emph{J.
  Opt. Soc. Am. A} \textbf{1994}, \emph{11}, 963--983\relax
\mciteBstWouldAddEndPuncttrue
\mciteSetBstMidEndSepPunct{\mcitedefaultmidpunct}
{\mcitedefaultendpunct}{\mcitedefaultseppunct}\relax
\EndOfBibitem
\bibitem[Feng \latin{et~al.}(2017)Feng, El-Ganainy, and Ge]{feng2017non}
Feng,~L.; El-Ganainy,~R.; Ge,~L. Non-Hermitian photonics based on parity--time
  symmetry. \emph{Nature Photonics} \textbf{2017}, \emph{11}, 752--762\relax
\mciteBstWouldAddEndPuncttrue
\mciteSetBstMidEndSepPunct{\mcitedefaultmidpunct}
{\mcitedefaultendpunct}{\mcitedefaultseppunct}\relax
\EndOfBibitem
\bibitem[Miri and Al\`u(2019)Miri, and Al\`u]{miri2019exceptional}
Miri,~M.-A.; Al\`u,~A. Exceptional points in optics and photonics.
  \emph{Science} \textbf{2019}, \emph{363}, eaar7709\relax
\mciteBstWouldAddEndPuncttrue
\mciteSetBstMidEndSepPunct{\mcitedefaultmidpunct}
{\mcitedefaultendpunct}{\mcitedefaultseppunct}\relax
\EndOfBibitem
\bibitem[Tuniz \latin{et~al.}(2022)Tuniz, Schmidt, and
  Kuhlmey]{tuniz2022influence}
Tuniz,~A.; Schmidt,~M.~A.; Kuhlmey,~B.~T. Influence of non-Hermitian mode
  topology on refractive index sensing with plasmonic waveguides.
  \emph{Photonics Research} \textbf{2022}, \emph{10}, 719--730\relax
\mciteBstWouldAddEndPuncttrue
\mciteSetBstMidEndSepPunct{\mcitedefaultmidpunct}
{\mcitedefaultendpunct}{\mcitedefaultseppunct}\relax
\EndOfBibitem
\bibitem[Bogaerts and Chrostowski(2018)Bogaerts, and
  Chrostowski]{bogaerts2018silicon}
Bogaerts,~W.; Chrostowski,~L. Silicon photonics circuit design: methods, tools
  and challenges. \emph{Laser \& Photonics Reviews} \textbf{2018}, \emph{12},
  1700237\relax
\mciteBstWouldAddEndPuncttrue
\mciteSetBstMidEndSepPunct{\mcitedefaultmidpunct}
{\mcitedefaultendpunct}{\mcitedefaultseppunct}\relax
\EndOfBibitem
\bibitem[Tuniz and Schmidt(2016)Tuniz, and Schmidt]{tuniz2016broadband}
Tuniz,~A.; Schmidt,~M.~A. Broadband efficient directional coupling to
  short-range plasmons: towards hybrid fiber nanotips. \emph{Optics Express}
  \textbf{2016}, \emph{24}, 7507--7524\relax
\mciteBstWouldAddEndPuncttrue
\mciteSetBstMidEndSepPunct{\mcitedefaultmidpunct}
{\mcitedefaultendpunct}{\mcitedefaultseppunct}\relax
\EndOfBibitem
\bibitem[Tuniz \latin{et~al.}(2019)Tuniz, Wieduwilt, and
  Schmidt]{tuniz2019tuning}
Tuniz,~A.; Wieduwilt,~T.; Schmidt,~M.~A. Tuning the Effective {PT} Phase of
  Plasmonic Eigenmodes. \emph{Physical Review Letters} \textbf{2019},
  \emph{123}, 213903\relax
\mciteBstWouldAddEndPuncttrue
\mciteSetBstMidEndSepPunct{\mcitedefaultmidpunct}
{\mcitedefaultendpunct}{\mcitedefaultseppunct}\relax
\EndOfBibitem
\bibitem[Akiki \latin{et~al.}(2020)Akiki, Verstuyft, Kuyken, Walter, Faucher,
  Lampin, Ducournau, and Vanwolleghem]{akiki2020high}
Akiki,~E.; Verstuyft,~M.; Kuyken,~B.; Walter,~B.; Faucher,~M.; Lampin,~J.-F.;
  Ducournau,~G.; Vanwolleghem,~M. High-Q THz photonic crystal cavity on a
  low-loss suspended silicon platform. \emph{IEEE Transactions on Terahertz
  Science and Technology} \textbf{2020}, \emph{11}, 42--53\relax
\mciteBstWouldAddEndPuncttrue
\mciteSetBstMidEndSepPunct{\mcitedefaultmidpunct}
{\mcitedefaultendpunct}{\mcitedefaultseppunct}\relax
\EndOfBibitem
\bibitem[Verstuyft \latin{et~al.}(2022)Verstuyft, Akiki, Vanwolleghem,
  Ducournau, Lampin, Walter, Bavedila, Lebouvier, Faucher, and
  Kuyken]{verstuyft2022short}
Verstuyft,~M.; Akiki,~E.; Vanwolleghem,~M.; Ducournau,~G.; Lampin,~J.-F.;
  Walter,~B.; Bavedila,~F.; Lebouvier,~{\'E}.; Faucher,~M.; Kuyken,~B. Short
  bends using curved mirrors in silicon waveguides for terahertz waves.
  \emph{Optics Express} \textbf{2022}, \emph{30}, 6656--6670\relax
\mciteBstWouldAddEndPuncttrue
\mciteSetBstMidEndSepPunct{\mcitedefaultmidpunct}
{\mcitedefaultendpunct}{\mcitedefaultseppunct}\relax
\EndOfBibitem
\bibitem[Jepsen \latin{et~al.}(2011)Jepsen, Cooke, and
  Koch]{jepsen2011terahertz}
Jepsen,~P.~U.; Cooke,~D.~G.; Koch,~M. Terahertz spectroscopy and
  imaging--Modern techniques and applications. \emph{Laser \& Photonics
  Reviews} \textbf{2011}, \emph{5}, 124--166\relax
\mciteBstWouldAddEndPuncttrue
\mciteSetBstMidEndSepPunct{\mcitedefaultmidpunct}
{\mcitedefaultendpunct}{\mcitedefaultseppunct}\relax
\EndOfBibitem
\bibitem[Caspers \latin{et~al.}(2013)Caspers, Aitchison, and
  Mojahedi]{caspers2013experimental}
Caspers,~J.~N.; Aitchison,~J.~S.; Mojahedi,~M. Experimental demonstration of an
  integrated hybrid plasmonic polarization rotator. \emph{Optics Letters}
  \textbf{2013}, \emph{38}, 4054--4057\relax
\mciteBstWouldAddEndPuncttrue
\mciteSetBstMidEndSepPunct{\mcitedefaultmidpunct}
{\mcitedefaultendpunct}{\mcitedefaultseppunct}\relax
\EndOfBibitem
\bibitem[Tuniz \latin{et~al.}(2020)Tuniz, Bickerton, Diaz, K{\"a}sebier, Kley,
  Kroker, Palomba, and de~Sterke]{tuniz2020modular}
Tuniz,~A.; Bickerton,~O.; Diaz,~F.~J.; K{\"a}sebier,~T.; Kley,~E.-B.;
  Kroker,~S.; Palomba,~S.; de~Sterke,~C.~M. Modular nonlinear hybrid plasmonic
  circuit. \emph{Nature Communications} \textbf{2020}, \emph{11}, 1--8\relax
\mciteBstWouldAddEndPuncttrue
\mciteSetBstMidEndSepPunct{\mcitedefaultmidpunct}
{\mcitedefaultendpunct}{\mcitedefaultseppunct}\relax
\EndOfBibitem
\bibitem[Kim and Qi(2015)Kim, and Qi]{kim2015polarization}
Kim,~S.; Qi,~M. Polarization rotation and coupling between silicon waveguide
  and hybrid plasmonic waveguide. \emph{Optics Express} \textbf{2015},
  \emph{23}, 9968--9978\relax
\mciteBstWouldAddEndPuncttrue
\mciteSetBstMidEndSepPunct{\mcitedefaultmidpunct}
{\mcitedefaultendpunct}{\mcitedefaultseppunct}\relax
\EndOfBibitem
\bibitem[Stefani \latin{et~al.}(2022)Stefani, Kuhlmey, Digweed, Davies, Ding,
  Zreiqat, Mirkhalaf, and Tuniz]{stefani2022flexible}
Stefani,~A.; Kuhlmey,~B.~T.; Digweed,~J.; Davies,~B.; Ding,~Z.; Zreiqat,~H.;
  Mirkhalaf,~M.; Tuniz,~A. Flexible terahertz photonic light-cage modules for
  in-core sensing and high temperature applications. \emph{ACS Photonics}
  \textbf{2022}, \emph{9}, 2128--2141\relax
\mciteBstWouldAddEndPuncttrue
\mciteSetBstMidEndSepPunct{\mcitedefaultmidpunct}
{\mcitedefaultendpunct}{\mcitedefaultseppunct}\relax
\EndOfBibitem
\bibitem[Rodrigo \latin{et~al.}(2015)Rodrigo, Limaj, Janner, Etezadi,
  Garc{\'\i}a~de Abajo, Pruneri, and Altug]{rodrigo2015mid}
Rodrigo,~D.; Limaj,~O.; Janner,~D.; Etezadi,~D.; Garc{\'\i}a~de Abajo,~F.~J.;
  Pruneri,~V.; Altug,~H. Mid-infrared plasmonic biosensing with graphene.
  \emph{Science} \textbf{2015}, \emph{349}, 165--168\relax
\mciteBstWouldAddEndPuncttrue
\mciteSetBstMidEndSepPunct{\mcitedefaultmidpunct}
{\mcitedefaultendpunct}{\mcitedefaultseppunct}\relax
\EndOfBibitem
\bibitem[Yang \latin{et~al.}(2021)Yang, Tang, Hu, Tang, Zhang, Cui, Wang,
  Chang, Fan, Li, \latin{et~al.} others]{yang2021near}
Yang,~Z.; Tang,~D.; Hu,~J.; Tang,~M.; Zhang,~M.; Cui,~H.-L.; Wang,~L.;
  Chang,~C.; Fan,~C.; Li,~J.; others Near-Field Nanoscopic Terahertz Imaging of
  Single Proteins. \emph{Small} \textbf{2021}, \emph{17}, 2005814\relax
\mciteBstWouldAddEndPuncttrue
\mciteSetBstMidEndSepPunct{\mcitedefaultmidpunct}
{\mcitedefaultendpunct}{\mcitedefaultseppunct}\relax
\EndOfBibitem
\bibitem[Tuniz(2021)]{tuniz2021nanoscale}
Tuniz,~A. Nanoscale nonlinear plasmonics in photonic waveguides and circuits.
  \emph{La Rivista del Nuovo Cimento} \textbf{2021}, \emph{44}, 193--249\relax
\mciteBstWouldAddEndPuncttrue
\mciteSetBstMidEndSepPunct{\mcitedefaultmidpunct}
{\mcitedefaultendpunct}{\mcitedefaultseppunct}\relax
\EndOfBibitem
\bibitem[Herter \latin{et~al.}(2023)Herter, Shams-Ansari, Settembrini, Warner,
  Faist, Lon{\v{c}}ar, and Benea-Chelmus]{herter2023terahertz}
Herter,~A.; Shams-Ansari,~A.; Settembrini,~F.~F.; Warner,~H.~K.; Faist,~J.;
  Lon{\v{c}}ar,~M.; Benea-Chelmus,~I.-C. Terahertz waveform synthesis in
  integrated thin-film lithium niobate platform. \emph{Nature Communications}
  \textbf{2023}, \emph{14}, 11\relax
\mciteBstWouldAddEndPuncttrue
\mciteSetBstMidEndSepPunct{\mcitedefaultmidpunct}
{\mcitedefaultendpunct}{\mcitedefaultseppunct}\relax
\EndOfBibitem
\bibitem[Guo \latin{et~al.}(2013)Guo, Ma, Wang, and Tong]{guo2013nanowire}
Guo,~X.; Ma,~Y.; Wang,~Y.; Tong,~L. Nanowire plasmonic waveguides, circuits and
  devices. \emph{Laser \& Photonics Reviews} \textbf{2013}, \emph{7},
  855--881\relax
\mciteBstWouldAddEndPuncttrue
\mciteSetBstMidEndSepPunct{\mcitedefaultmidpunct}
{\mcitedefaultendpunct}{\mcitedefaultseppunct}\relax
\EndOfBibitem
\bibitem[Gacemi \latin{et~al.}(2013)Gacemi, Mangeney, Colombelli, and
  Degiron]{gacemi2013subwavelength}
Gacemi,~D.; Mangeney,~J.; Colombelli,~R.; Degiron,~A. Subwavelength metallic
  waveguides as a tool for extreme confinement of THz surface waves.
  \emph{Scientific Reports} \textbf{2013}, \emph{3}, 1369\relax
\mciteBstWouldAddEndPuncttrue
\mciteSetBstMidEndSepPunct{\mcitedefaultmidpunct}
{\mcitedefaultendpunct}{\mcitedefaultseppunct}\relax
\EndOfBibitem
\bibitem[Br{\"u}ckner \latin{et~al.}(2009)Br{\"u}ckner, K{\"a}sebier,
  Pradarutti, Riehemann, Notni, Kley, and
  T{\"u}nnermann]{bruckner2009broadband}
Br{\"u}ckner,~C.; K{\"a}sebier,~T.; Pradarutti,~B.; Riehemann,~S.; Notni,~G.;
  Kley,~E.-B.; T{\"u}nnermann,~A. Broadband antireflective structures applied
  to high resistive float zone silicon in the {THz} spectral range.
  \emph{Optics Express} \textbf{2009}, \emph{17}, 3063--3077\relax
\mciteBstWouldAddEndPuncttrue
\mciteSetBstMidEndSepPunct{\mcitedefaultmidpunct}
{\mcitedefaultendpunct}{\mcitedefaultseppunct}\relax
\EndOfBibitem
\end{mcitethebibliography}

\end{document}